\documentclass[aip,jcp,
11pt,
amsmath,amssymb,
floatfix,
reprint,
]{revtex4-1}

\usepackage[utf8]{inputenc} 
\usepackage[T1]{fontenc} 
\usepackage{hyperref}
\usepackage{verbatim}

\usepackage[table]{xcolor} 
\usepackage{graphicx}

\usepackage{dcolumn}%
\usepackage{bm}%

\usepackage{listings}

\usepackage[version=3]{mhchem}
\usepackage{units}
\usepackage{siunitx}

\usepackage{xspace}

\newcommand{\Ref}{Ref.\xspace}
\newcommand{\Refs}{Refs.\xspace}
\newcommand{\lit}[1]{\Ref\citenum{#1}\xspace}
\newcommand{\lits}[1]{\Refs\citenum{#1}\xspace}

\newcommand{\braket}[2]{\ensuremath{ \langle #1 | \, #2  \rangle }}

\newcommand{\ket}[1]{\ensuremath{  | {#1} \rangle}}
\newcommand{\bra}[1]{\ensuremath{\langle {#1} | }}
\newcommand{\Ket}[1]{\ensuremath{\left| {#1} \right \rangle }}

\newcommand{\partdd}[2]{\ensuremath{ \frac{\partial^2 {#1}}
{\partial {#2}^2} }}

\renewcommand{\i}{\ensuremath{\mathrm{i}}}
\newcommand{\ii}{\ensuremath{\mathrm{i}}}

\newcommand{\dd}{\ensuremath{\mathrm{d}}}

\definecolor{ocre}{RGB}{243,102,25}
\definecolor{mygray}{RGB}{243,243,244}
\definecolor{fzjred}{RGB}{175,90,80}

\definecolor{blau}{HTML}{1F78B4}
\definecolor{gruen}{HTML}{33A02C}
\definecolor{hellblau}{HTML}{A6CEE3}
\definecolor{hellgruen}{HTML}{B2DF8A}

\definecolor{nrot}{HTML}{d7191c}
\definecolor{norange}{RGB}{253,174,97}
\definecolor{ngruen}{HTML}{abdda4}
\definecolor{nblau}{HTML}{2b83ba}

\definecolor{nrot1}{RGB}{215,48,31}
\definecolor{nrot2}{RGB}{252,141,89}
\definecolor{nrot3}{RGB}{253,204,138}
\definecolor{nrot4}{RGB}{254,240,217}

\definecolor{CBred}{RGB}{215,25,28}
\definecolor{CBorange}{RGB}{253,174,97}
\definecolor{CByellow}{RGB}{255,255,191}
\definecolor{CBgreen}{RGB}{171,211,164}
\definecolor{CBlgreen}{RGB}{166,217,106}
\definecolor{CBdgreen}{RGB}{26,150,65}
\definecolor{CBblue}{RGB}{43,131,186}
\definecolor{CBblue2}{RGB}{146,197,222}
\definecolor{CBdblue}{RGB}{5,113,176}
\definecolor{CBgray60}{RGB}{102,102,102}
\definecolor{CBgray20}{RGB}{204,204,204}

\let\oldtheequation\theequation
\makeatletter
\def\tagform@#1{\maketag@@@{\ignorespaces#1\unskip\@@italiccorr}}
\renewcommand{\theequation}{(\oldtheequation)}
\makeatother 
\begin{document}
  \title{Resonance dynamics of \ce{DCO} ($\widetilde{X}\,{}^2A'$) simulated with 
the
dynamically pruned discrete variable representation (DP-DVR)}

\author{Henrik R.~Larsson}
\email{larsson@pctc.uni-kiel.de}
\author{Jens Riedel}
\altaffiliation[Present address: ]
{Federal Institute for Materials Research and Testing (BAM), Unter den Eichen 
87, 12205 Berlin, Germany}
\author{Jie Wei}
\author{Friedrich Temps}
\author{Bernd Hartke}
\affiliation{Institut für Physikalische Chemie, Christian-Albrechts-Universität zu Kiel, Olshausenstraße 40, 24098 Kiel, Germany}
\date{\today}
%
%
%
%
%
%
%
\begin{abstract}
  Selected resonance {states} of the deuterated formyl radical in the 
electronic ground state 
$\widetilde{X}\,{}^2A'$ are computed using our recently introduced 
dynamically pruned discrete variable representation (DP-DVR) [H.~R.~Larsson, B.~Hartke 
and 
D.~J.~Tannor, \textit{J.~Chem.~Phys.}, \textbf{145}, 204108 (2016)].
Their decay and asymptotic distributions are analyzed and, for selected resonances, 
compared to experimental results obtained by a combination of stimulated 
emission pumping (SEP) and velocity-map imaging of the product \ce{D} 
atoms. The theoretical results show good 
agreement with the experimental kinetic energy distributions. 
The intramolecular vibrational energy redistribution (IVR) is analyzed and compared with 
previous 
results from an effective polyad Hamiltonian. 
{Specifically, we analyzed the part of the wavefunction that remains in the 
interaction  region during the decay.}
{The} results {from the polyad Hamiltonian} could mainly be confirmed.
The \ce{C=O} stretch quantum number is typically conserved, while 
the \ce{D-C=O} bend quantum number decreases.
Differences are due to strong anharmonic coupling such that all resonances have 
major contributions from several zero-order states. For some of the resonances, the 
coupling is so strong that no further zero-order states appear during 
the dynamics {in the interaction region}, even after propagating for 
\unit[300]{ps}.
\end{abstract}

\maketitle

\clearpage
\section{Introduction}
\label{sec:intro}
One of the fundamental processes in molecular reaction dynamics is the unimolecular 
dissociation of vibrationally excited 
molecules.%
\cite{robinson_holbrook_book,baer_hase_book,schinke_book,stumpf_1995}
Depending on the molecular system, their decay dynamics can span the range from 
mode-specific to statistical.
For intermediate cases, the dissociation mechanism
in terms of intramolecular 
vibrational energy redistribution (IVR) 
of bound and metastable resonance states provides considerable 
insight into the dynamic properties of the molecular system under study.
{In the following, we use the term ``resonance'' both for vibrational 
resonances and metastable states. The meaning should be clear from the 
context and typically it will mean metastable state. Otherwise, we will point 
out the meaning.} 

One standard benchmark system is the formyl radical \ce{HCO} which plays an important 
role in many combustion processes as well as in atmospheric and interstellar 
chemistry; see, e.g., \lit{hco_astro_ocana_2017}.
In its electronic ground state $\widetilde{X}\,{}^2A'$, \ce{HCO} shows many 
resonance states whose decays follow very systematic mode-specific pathways.

In contrast to \ce{HCO}, its deuterated counterpart \ce{DCO} shows more 
statistical behavior. 
This is due to an accidental, strong 
$\nu_1\!:\!\nu_2\!:\!\nu_3\!\approx\!1\!:\!1\!:\!2$ Fermi 
{(vibrational)} resonance in 
the \ce{D-CO}  stretching  vibration $\nu_1$ ($\unit[1910]{cm^{-1}}$), the \ce{DC=O} stretching vibration $\nu_2$ ($\unit[1795]{cm^{-1}}$), and the \ce{D-C=O} bending vibration $\nu_3$ ($\unit[847]{cm^{-1}}$). 
The Fermi resonance leads to strong mixing between zero-order states, 
already for the lowest vibrational states.\cite{DCO_I_schinke,DCO_II_schinke,DCO_polyad_troellsch_2001,Jung2002} 
Strong mixing is an essential prerequisite for 
statistical dynamics. Keller \emph{et al.} thus consider \ce{DCO} a ``precursor of 
`chaos' 
in more complicated systems''.\cite{DCO_II_schinke}
Consequentially, the normal assignments in terms of the associated vibrational quantum 
numbers ($v_1$, $v_2$, $v_3$) loose their meanings, although they may still be used 
economically as ``nominal labels'' to indicate possible predominating vibrational 
character. The only conserved quantity at short times is the polyad quantum 
number, $P = v_1 + v_2 +v_3/2$, which describes the total vibrational excitation.%
\cite{polyad_quack_1984,polyad_quack_1985,polyad_rev_kellman_2007,DCO_I_schinke} 

Due to their strikingly different characteristics, there has been strong interest in both \ce{HCO} 
and \ce{DCO}, from both experiment and theory.
Due to the amount of work on these systems, we briefly mention only some
of the most important contributions.
Energies and widths of resonances of \ce{HCO} and \ce{DCO} 
have 
been measured to high 
resolution by dispersed fluorescence and stimulated emission 
pumping.%
\cite{HCO_exp_adamson_1993,HCO_exp_neyer_1993,HCO_exp_tobiason_1995, 
DCO_exp_tobiason_1995,DCO_I_schinke,DCO_wei_2004}
For \ce{HCO}, Neyer \emph{et al.} also measured the rovibrational product distribution 
of \ce{CO}.\cite{HCO_exp_neyer_1995}

Likewise, energies and widths of \ce{HCO} have been computed using 
various methods, establishing this system as a benchmark for computing 
resonance states.%
\cite{HCO_bowman_1986a,HCO_romanowski_1986,HCO_I_schinke,HCO_II_schinke,HCO_IV_schinke,
HCO_poirier_2002a, HCO_mandelshtam_2002, 
HCO_tremblay_2005,HCO_mctdh_ndengue_2015,HCO_pes_ndengue_2016} 
Rovibrational \ce{CO} state distributions and resonance decays have been 
analyzed by various researchers, also for 
nonzero total angular 
momentum.\cite{HCO_qi_1996,HCO_yang_1997,HCO_I_schinke,HCO_II_schinke,HCO_III_schinke,
HCO_V_schinke,dixon_1992,gray_1992} 
However, less attention has been paid to 
\ce{DCO}.\cite{DCO_wang_1995,DCO_I_schinke,DCO_II_schinke,DCO_loettgers_1997,
stamatiadis_2001}
Although \ce{DCO} shows close to statistical behavior, it is not fully irregular. 
The decay constants show strong state-to-state fluctuations. Microcanonical 
statistical rate theory thus cannot predict the state-specific decay 
constants.
Due to strong anharmonic couplings, an IVR analysis is much less straightforward but 
still desirable.
Clear indications that remaining regularities may be strong enough for such an
approach were provided by the aforementioned %
polyad model 
that turned out 
to be applicable to \ce{DCO} and quite useful.\cite{DCO_II_schinke}
The group of Temps fitted an effective polyad Hamiltonian to experimental data and did an 
analysis of different IVR pathways within this polyad
model.\cite{DCO_polyad_troellsch_2001,DCO_polyad_renth_2003}
This model Hamiltonian was then used for further semiclassical 
analysis.\cite{jung_2002,huang_2007,DCO_semparithi_2003}
However, many of these analyses still lack comparison with quantum calculations on an 
\emph{ab-initio} potential energy surface (PES). 

Here, we make an initial attempt to fill this gap by a joint experimental and theoretical 
study of the observed and calculated resonances of \ce{DCO} in polyads $5$ and $5.5$. In 
particular, we present {experimental} kinetic energy 
release (KER) 
spectra of the \ce{D} atoms from the decay of the resonances and the associated \ce{CO} 
vibration-rotation product state distributions. The results were obtained using the method 
of Stimulated Emission Pumping (SEP)\cite{DCO_I_schinke} for the preparation of selected 
excited \ce{DCO} resonance states in combination with velocity-map imaging 
\cite{Eppink1997} of 
the \ce{D} atoms. In addition, we computed the \ce{CO} state distributions using our 
dynamically 
pruned discrete variable representation (DP-DVR) 
approach\cite{proDG_hartke_2006,pW_tannor_2016,pvb_rev_tannor_2017} and analyzed 
the computed resonance decays. As critical tests of the theory, we compare our 
computational results with the experimental data taking either the PES of Werner 
\emph{et al.} (WKS)\cite{HCO_I_schinke,HCO_II_schinke}
or the PES of Song \emph{et al.} (SAG).\cite{HCO_pes_song_2013}

We thereby use this 
study as a real-life 
application of our new 
method for performing molecular quantum dynamics.
The standard method in molecular quantum dynamics is to employ a direct-product basis of 
discrete variable representation (DVR) functions.\cite{tannor_book}
However, this approach suffers from an exponential scaling of basis size with 
dimensionality. Even for lower-dimensional problems like the one studied here, it is not 
the most efficient method. For dissociation problems like \ce{DCO}, long 
coordinate ranges need to be described, but the wavefunction typically does not 
occupy the whole direct-product coordinate space. Such a problem is well-suited for 
only using DVR grid points where the wavepacket has non-negligible amplitudes. This leads 
to significant savings in computational resources (both regarding number of operations 
and memory requirements). Since the wavepacket moves in time, the set of active basis 
functions needs to be dynamically adapted. Exactly this capability is provided by our 
dynamically pruned DVR 
approach (DP-DVR).\cite{proDG_hartke_2006,pW_tannor_2016,pvb_rev_tannor_2017}
The accuracy is controlled by a so-called wave-amplitude 
threshold.\cite{proDG_hartke_2006,pvb_algorithms_tannor_2016} The smaller the threshold, 
the more basis functions are added and the more accurate the simulation becomes.
In \lit{pvb_rev_tannor_2017} we have introduced a very efficient algorithm for DP 
simulations and have done careful benchmarks of the DP-DVR and other methods. 
We have shown that DP-DVR can be more efficient than conventional dynamics
already for two-dimensional systems.

For the dissociation dynamics of \ce{DCO}, phase-space bases would be other useful 
candidates 
for our DP 
approach.\cite{pvb_tannor_2012,pvb_math_tannor_2016,pW_tannor_2016,pvb_rev_tannor_2017} 
One could use such a phase-space basis in the coordinate describing the dissociation. 
However, efficient use of phase-space bases requires that the potential operator has the 
form of a sum of products of one-dimensional terms (SOP form). This requires an 
additional fitting of the potential into this form. 

Recently, we have combined our DP approach with the Multiconfiguration 
Time-De\-pen\-dent Hartree method (MCTDH), giving DP-MCTDH.\cite{dpmctdh_tannor_2017}
There, DP can be used either for pruning the time-dependent basis functions 
(single-particle functions) or their DVR representation. The former is most useful for 
higher-dimensional systems (see also 
\lits{mctdh_selected_configurations_worth_2000,pruned_mctdh_carrington_2016,
pruned_mctdh_carrington_2017}) whereas the latter is useful whenever many DVR functions 
are needed for describing the (multidimensional) single-particle functions.
This is the case for \ce{DCO} dissociation dynamics such that DP-MCTDH could be a useful 
method for our study. MCTDH itself has been used for computing resonances of 
\ce{HCO}.\cite{HCO_mctdh_ndengue_2015,HCO_pes_ndengue_2016}
However, the MCTDH algorithm is more complicated than a DVR code. Furthermore,
as with phase-space bases, an efficient use of MCTDH requires a SOP form of the potential.
Although methods for combining arbitrary potentials with MCTDH 
exist,\cite{cdvr_manthe_1996,CMCTDH_wodraszka_2018} they are 
less well established and require more careful convergence tests. 
Further, MCTDH is most useful for weakly correlated systems and short-time dynamics, but 
for \ce{HCO} \cite{HCO_mctdh_ndengue_2015} many single-particle functions are needed and 
the wavepacket has to be propagated for tens of picoseconds.

Here, we resort to our standard DP-DVR algorithm to make our simulations as simple as 
possible, without jeopardizing computational 
efficiency, and also to test 
the capabilities of our DP-DVR approach in an application involving resonances
and their decay.

The remainder of this Article is organized as follows: 
The experimental and theoretical setups are described in Sections \ref{sec:exp} and 
\ref{sec:theory}, respectively. The latter provides more details on the DP-DVR method 
(\autoref{sec:dp-dvr}), on the methods to obtain resonance states 
(\autoref{sec:filter_diag}) and asymptotic product distributions 
(\autoref{sec:theory_rovib}), and on the employed PES (\autoref{sec:theory_pes}).
Our results are presented and discussed in \autoref{sec:results}. 
The experimental velocity map images are presented in \autoref{sec:res_exp}.
\autoref{sec:res_params} details the simulation parameters and
\autoref{sec:energies} compares our computed resonance energies and widths with literature 
values. The PES is analyzed in 
terms of an adiabatic picture in \autoref{sec:adiabatic_repr}, the experimental and 
theoretical asymptotic distributions are presented and compared in 
\autoref{sec:rovib_distr}, and the decay processes of the studied resonances
are analyzed in 
\autoref{sec:decay_processes}. We summarize in \autoref{sec:conclusion}.

\section{Experiment}
\label{sec:exp}
\begin{figure*}
        \includegraphics[width=.8\textwidth]{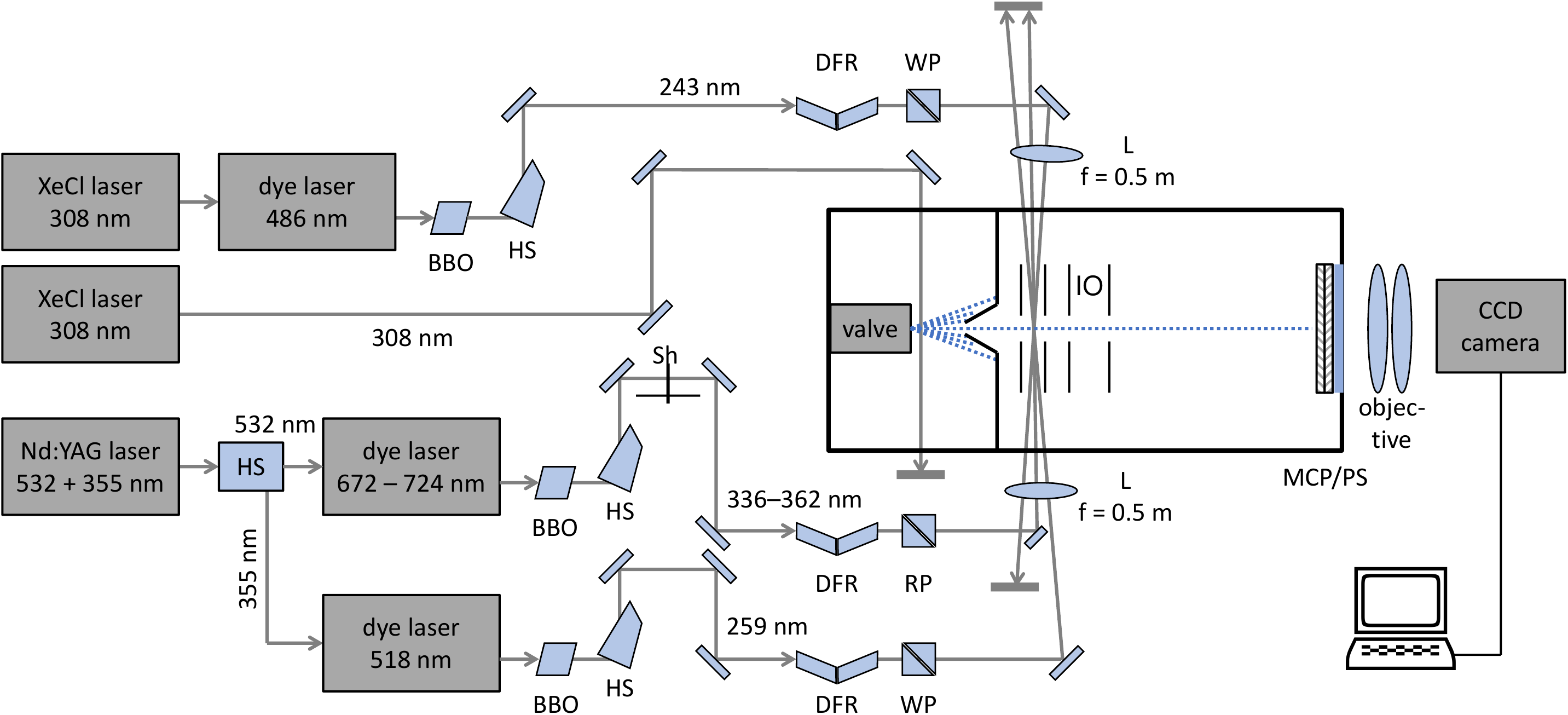}
        \caption{%
Schematic diagram of the experimental setup. BBO: doubling crystal, DFR: 
double-Fresnel rhomb, HS: harmonic separator, IO: ion optics electrode assembly, L: lens, 
MCP: multi-channel plate, PS: phosphor screen, RP: Rochon prism, Sh: shutter, WP: 
Wollaston prism.
        }
        \label{fig:setup}
\end{figure*}

The {measurements} required four spatially and temporally controlled laser 
pulses to be focused into a supersonic seeded molecular beam in a differentially pumped 
stainless steel vacuum chamber. 
\autoref{fig:setup} shows a sketch of the employed installation. 
The SEP part of the setup for preparation of the \ce{DCO} ($ \widetilde X$) radicals 
in 
their 
highly excited states \cite{DCO_I_schinke,DCO_wei_2004} and the photofragment imaging part 
for measuring the kinetic energy release to the \ce{D} 
atoms\cite{Wei2003,HCO_riedel_2005} 
have been described in some detail separately before. 
The exact experimental conditions varied slightly from those to record optimal SEP 
spectra by the need of the present experiment for a high number density of highly 
vibrationally excited radicals.\cite{riedel_thesis}

A pulsed supersonic free jet containing $\sim \unit[0.5]{\%}$ deuterated acetaldehyde 
(\ce{CD3CDO}) in \ce{He} was generated by flowing the carrier gas at a backing 
pressure of 
$\sim \unit[2]{bar}$ through a stainless steel reservoir with freshly distilled 
\ce{CD3CDO}
(Fluka, $>\unit[99]{\%}$) at $\unit[-78]{^\circ{}C}$ and expanding it into the 
first vacuum chamber of 
the imaging apparatus through a 0.8 mm diameter solenoid-actuated valve (General Valve) 
at \unit[20]{Hz} repetition rate. 
Photolysis of the \ce{CD3CDO} with a \ce{XeCl} excimer laser at $\lambda = \unit[308]{nm}$
in the 
high-pressure region of the free jet expansion $\sim \unit[3]{mm}$ behind the 
pulsed nozzle 
produced the desired \ce{DCO} radicals.
A molecular beam of the radicals shaped by a \unit[1.5]{mm} diameter conical skimmer then 
entered 
the test volume between the repeller and extractor plates of the imaging electrode 
assembly in the second vacuum chamber, where it was crossed by the pump, dump and probe 
laser beams. A liquid-\ce{N2} cryo-pump surrounding the assembly minimized the 
background 
ion signal. 

Two dye lasers (Lambda Physik) were optically pumped by the dichroically separated third 
resp.~second harmonic output beams of a Nd:YAG laser (Spectra Physics) and 
frequency-doubled in BBO crystals to obtain the required $\sim\unit[259]{nm}$ 
pump (\unit[1.5]{mJ}) and 
\unit[336\,--\,362]{nm} dump (\unit[5\,--\,7]{mJ}) pulses. 
The precise wavelengths were set by recording the $ \widetilde B (^2A')\leftarrow 
\widetilde X(^2A')$ 
fluorescence excitation and $ \widetilde B \rightarrow \widetilde X$ SEP spectra in a 
separate molecular beam apparatus with fluorescence detection. 
\cite{DCO_I_schinke,DCO_wei_2004} 
Both beams were focused into the test volume at a small angle with respect to each other 
through a $f=\unit[500]{mm}$ fused silica lens.
The \unit[308]{nm} excimer-pumped probe dye laser (both Lambda Physik) for imaging of the 
\ce{D} atoms 
from the \ce{D-CO} dissociation reaction by 2+1 resonance-enhanced multi-photon 
ionization 
(REMPI) via the $2\,{}^2S \leftarrow 1\,{}^2S$ transition was frequency-doubled to the 
required $\lambda = \unit[243]{nm}\ (\unit[1]{mJ})$ and focused into the 
detection region 
counter-propagating to the pump and dump with a short delay of $\sim\unit[10]{ns}$.
The exact probe wavelength was set by optimizing the \ce{D^+} ion yield. 
All laser polarizations were set parallel to the plane of the imaging detector using 
combinations of Fresnel double rhombs and Rochon or Wollaston prisms. 

The imaging measurements were made under velocity mapping conditions. \cite{Eppink1997} 
The probe laser was periodically scanned over the Doppler profile of the recoiling \ce{D}
atoms.  
The resulting \ce{D^+} ions were monitored on a microchannel plate (MCP) detector 
coupled to 
a phosphor screen. 
The obtained signals were recorded on a CCD camera and accumulated over up to
$200\, 000$ laser shots using single ion counting and centroiding \cite{Chang1998} to 
improve 
the detection sensitivity
and spatial resolution and to discriminate against noise. 
Background images with the dump laser blocked on alternate shots were subtracted to 
eliminate contributions 
to the \ce{D} atoms from the \ce{CD3CDO} photolysis and from predissociation of the 
\ce{DCO}
($\widetilde B$) state.
Mass selectivity was achieved using a fast transistor switch (Behlke) to gate the MCP.
The timing of the pulsed valve, all lasers and the MCP voltage was controlled by a 
digital delay generator (Stanford Research).

\section{Theory}
\label{sec:theory}
\subsection{Dynamically pruned discrete variable representation (DP-DVR)}
\label{sec:dp-dvr}
The propagation is performed in Jacobi coordinates $\{R,r,\theta\}$. $R$ is the distance 
of \ce{D} to the center of mass of \ce{C-O}, $r$ is the \ce{C-O} distance and $\theta$ 
the angle between the corresponding vectors $\vec{R}$ and $\vec{r}$. 
As usual,\cite{jacobi_coords_lequere_1990} the wavefunction is 
divided by $R r$ such that the volume element takes the form of $\dd V = \dd R\, \dd r\, 
\sin(\theta) \dd \theta$.
In our propagation, we consider only the $\widetilde{X}\,{}^2A'$ ground state and neglect 
nonadiabatic and Renner-Teller couplings. The couplings only play an important role
if the wavefunction has non-negligible contributions at linear 
geometries.\cite{HCO_I_schinke} This requires high excitation in the bending mode and 
only occurs for a minority of the resonances studied; we will mention them in 
\autoref{sec:decay_processes}.

The wavefunction is represented by a Gauß-Legendre DVR 
for the angular coordinate and by a sinc DVR for the radial 
coordinates.\cite{tannor_book,sinc_dvr_miller_1992}
This direct-product basis is then pruned using our DP-DVR
approach,\cite{pW_tannor_2016} 
{which is related to previous pruning 
methods.\cite{proDG_hartke_2006,proDG_hartke_2008,pruning_wyatt_2006,
  pruning_wyatt_2007,pruning_mccormack_2006}
In contrast to those, 
we could show that our approach is actually faster than conventional methods.}
If the absolute value of a coefficient of a
direct-product basis function is larger than a predefined wave-amplitude threshold, this 
basis function and its nearest neighbors become active. Otherwise, this function 
is removed from the set of active functions. This procedure is repeated at each time 
step, ensuring a compact representation of the wavefunction for all propagation times. 
For further details we refer to \lit{pW_tannor_2016}. 

In the following, we mention two improvements of our code introduced in 
\lit{pW_tannor_2016} (the reader who is not interested in technical details may skip the 
rest of this Subsection).
The first improvement is a (straightforward) implementation of standard shared-memory 
parallelization. For the second improvement, 
we take advantage of the fact that the momentum and kinetic energy operators in 
sinc DVR representation give Toeplitz matrices with elements $T_{i,j} = T_{i+1,j+1} = 
t_{i-j}$. Instead of storing all matrix elements $T_{i,j}$, we 
only store one row $t_{i-j}$ of the matrix. This reduces the memory needed to be loaded 
into the cache of the central processing units and thus gives large speed-ups, especially 
for shared-memory parallelization. In standard molecular quantum dynamics 
applications with basis 
sizes smaller than 100, the storage of the one-dimensional matrices is negligible, 
compared to the storage of the (pruned) multidimensional wavefunction. However, the 
exploitation of the Toeplitz structure becomes useful whenever the one-dimensional basis 
size is large. This is the case for the dissociation dynamics considered here.
For a basis that is not pruned, the action of a Toeplitz matrix onto a vector can be 
implemented with a scaling of $\mathcal 
O\bigl(N\log(N)\bigr)$ using fast Fourier transforms 
(FFTs).\cite{toeplitz_nlogn_gohberg_1994} However, when the basis is pruned, some 
structure 
of the pruned matrix is lost and an implementation in terms of FFTs is not 
straightforward. 
Nevertheless, the matrix--vector product in a pruned basis is much faster than 
a FFT in an unpruned basis if pruning reduces the size of the objects
sufficiently, which typically is the case.

\subsection{Retrieval of resonances}
\label{sec:filter_diag}
Previously, \ce{HCO} and \ce{DCO} resonances have been computed using, among others, 
Lanczos procedures,\cite{HCO_poirier_2002a,HCO_tremblay_2005}
a log-derivative version of Kohn’s variational 
principle,\cite{DCO_II_schinke,HCO_II_schinke}
MCTDH,\cite{HCO_mctdh_ndengue_2015}
projection theory,\cite{HCO_wang_2017}
and filter diagonalization.\cite{HCO_mandelshtam_2002,HCO_V_schinke}
Many applications utilize the time-independent Schrödinger equation (TISE).
In principle, our DP-DVR code would work as well for the TISE using an algorithm that 
iteratively adds and removes new basis 
functions after each Lanczos iteration.\cite{pvb_algorithms_tannor_2016}
However, Lanczos and other iterative diagonalization algorithms require a good 
preconditioner for efficient 
usage,\cite{precond_huang_2000,precond_poirier_2001,HCO_poirier_2002a} especially when 
eigenstates are searched in the middle of a dense eigenspectrum, as is the case in 
this application. The search for a good preconditioner for a Hamiltonian is not trivial 
and depends on the employed basis. We tested the 
preconditioner proposed in \lit{precond_huang_2000} and some standard preconditioners 
like 
diagonal matrices but found no satisfying results. 

Thus, instead of using a Lanczos approach, we use filter 
diagonalization.\cite{filterdiag_neuhauser_1990,filterdiag_neuhauser_1991}
Since the standard form of filter diagonalization is based on propagation, it allows for 
a straightforward integration into a DP code for solving the time-dependent Schrödinger 
equation. 
This worked without any adjustments or tuning of parameters, and the 
bottleneck of our simulations was then not the retrieval of the eigenstates but the 
subsequential simulation of their decay.
Thus, we did not try to use further improvements of filter 
diagonalization.\cite{filterdiag_takatsuka_1995,filterdiag_taylor_1997,
HCO_mandelshtam_2002}

In standard filter diagonalization, a real-time dynamics of an initial wavepacket 
with absorbing boundary conditions is performed and the wavepacket 
is stored at intermediate times. Afterwards, the wavepacket is filtered at 
(typically equidistantly spaced) energies $E_n$ \emph{via} Fourier 
transform:\cite{filterdiag_neuhauser_1991}
\begin{equation}
 \ket{\Phi_F(E_n)} = \frac1{2\pi\hbar} \int_{0}^\infty \exp(\ii E_n t/\hbar) 
\ket{\Psi(t)} F(t) 
\dd t,
\end{equation}
where $F(t)$ is a filter function of the form
\begin{equation}
F(t) =   \exp[-(t-\tau)^2 W^2 / \hbar^2]. \label{eq:filterfunction}
\end{equation}
$W$ is the energy bandwidth, here taken as $E_{n+1}-E_{n}$.
The duration $\tau$ is set such that $W\tau \gg 1$.
Afterwards, the Hamiltonian is represented in the nonorthogonal basis 
of filtered states $\{\ket{\Phi_F(E_n)}\}_{n=1}^N$ and diagonalized. The size of the 
basis is typically less than $100$, such that diagonalization can be performed using 
standard procedures. 
{The obtained eigenvalues $E_n$ are complex-valued, including both the
resonance energies $\epsilon_n$ and resonance widths $\Gamma_n$ as 
$E_n = \epsilon_n - \ii \frac{\Gamma_n}{2}$.
}
{In the following, the width $\Gamma$ will be used for wavenumber quantities.}

\subsection{Rovibrational product distribution}
\label{sec:theory_rovib}

For obtaining the KER spectra of the \ce{D} atom and the asymptotic 
rovibrational distributions of the \ce{CO} fragment, we employ the analysis-line method 
developed by Balint-Kurti \emph{et 
al}.\cite{analysis_line_balint-kurti_1990} %
It has previously been used for \ce{HCO} by both Gray and 
Dixon.\cite{gray_1992,HCO_yang_1997,dixon_1992}
The retrieval of the asymptotic distribution is performed by Fourier-transforming 
cuts of $\Psi(R,r,\theta; t)$ along the ``analysis line'' $R=R_\infty$, where the cuts 
$\Psi(R_\infty, r,\theta;t)$ are represented in the basis of asymptotic rovibrational 
states, $\{\ket{\psi_{vj}}\}$. $v$ and  $j$ are the \ce{CO} stretch and 
rotational quantum numbers, respectively. The working expression is
\begin{align*}
C_{vj}(R_\infty,t) &= \bra{R_\infty}\braket{\psi_{vj}}{\Psi(t)},\\
A_{vj}(R_\infty,E) &= \frac1{2\pi} \int_0^\infty \exp(\ii E t/\hbar) 
C_{vj}(R_\infty,t)\dd t. 
\end{align*}
The KER spectrum $P(E_{\ce{D}})$ is then given by
\begin{align}
P(E_{\ce{D}}) &= \sum_{v,j=0}^{\infty} 
P_{vj}(E_{\ce{D}}),\label{eq:P_Ed_analysis_line}\\
P_{vj}(E_{R}) &=
\lim_{t\to\infty}\bigl| \braket{\Psi^{(-)}_{kvj}}{\Psi(t)}\bigr|^2 %
\label{eq:P_Ed_analysis_line_vj}\\
&= \frac{16 \pi^3 k}{\mu_R} \left| A_{vj}(R_\infty,E_R)\right|^2, \\
E &= E_{vj} + E_R, \quad E_D = \frac{m_{\ce{CO}}}{m_{\ce{DCO}}}  
E_R,\label{eq:ED_mass_weight}
\end{align}
where $k = \sqrt{2\mu_R E_{R}/\hbar^2}$ and  $E_R$ is the kinetic energy in the
internal Jacobi coordinate $R$. The corresponding kinetic energy of the \ce{D} atom 
in the laboratory frame (without translation and rotation of the \ce{DCO} system) 
is then given by mass-weighting $E_R$ (\autoref{eq:ED_mass_weight}), as obtained from the
standard center-of-mass transformation of a two-body system.\cite{bernath_book}
$E_{vj}$ is the internal energy of the \ce{CO} fragment. 
$\mu_R$ is the reduced mass of $\ce{D-CO}$ in Jacobi 
coordinates. $\ket{\Psi^{(-)}_{kvj}}$ is the outgoing scattering state for 
dissociation into \ce{D} and \ce{CO}.\cite{schinke_book}
Integrating $P_{vj}(E_{R})$ over all $E_{R}$ gives the rovibrational 
distributions of the \ce{CO} product.

Combining this method with our DP-DVR approach is straightforward because the only 
quantity that needs to be stored during the dynamics calculation is the wavefunction 
evaluated at 
$R_\infty$ which is a DVR grid point in the asymptote. When no basis function at 
$R_\infty$ 
is active at a specific time step, the wavefunction there is simply zero.

\subsection{Potential energy surfaces}
\label{sec:theory_pes}

To the best of our knowledge, there are three accurate and more recent potential energy 
surfaces (PES) for \ce{HCO}. The (modified) WKS surface of Werner, Keller and 
Schinke\cite{HCO_I_schinke,HCO_II_schinke} has been applied in many 
studies; for example %
\lits{DCO_I_schinke,DCO_II_schinke,HCO_poirier_2002a,HCO_mandelshtam_2002,%
HCO_tremblay_2005,HCO_mctdh_ndengue_2015,HCO_wang_2017}. 
This PES quite accurately describes both the dissociation and the interaction region, 
including the 
conical intersection at linear geometry. It is based on 
multireference configuration interaction (MRCI) calculations.
Another recent PES is that developed by Song, 
van der Avoird and Groeneboom (SAG).\cite{HCO_pes_song_2013} It is based on unrestricted 
coupled-cluster calculations and focuses on the asymptotic and lower-energy
regions. 
It has been used for scattering calculations.\cite{HCO_Scatt_song_2015}
The third recent PES was developed by Ndengu{\'e}, Dawes and Guo 
and is based on explicitly correlated MRCI calculations.\cite{HCO_pes_ndengue_2016}
It was used for unraveling the effects of Renner-Teller coupling on the resonance 
levels. 

Only the WKS surface has been used in studies of 
\ce{DCO}.\cite{DCO_I_schinke,DCO_II_schinke} To connect to those previous studies, we 
mainly use this well-established PES in our calculations. In order to estimate the 
sensitivity of the observables to changes of the potential, we also employed the SAG 
surface for selected resonances.
Figures i and ii in the supplementary material show cuts of the two PES.
Since none of those PES were constructed with decay dynamics of
high-energy resonance {states} far out into the asymptote
in mind, none of the utilized 
PES can be expected to give quantitative agreement between theoretical 
and experimental results in this present application.

\section{Results and discussion}
\label{sec:results}

\subsection{Experimental Results}
\label{sec:res_exp}

\begin{figure}
        \includegraphics[width=\columnwidth]{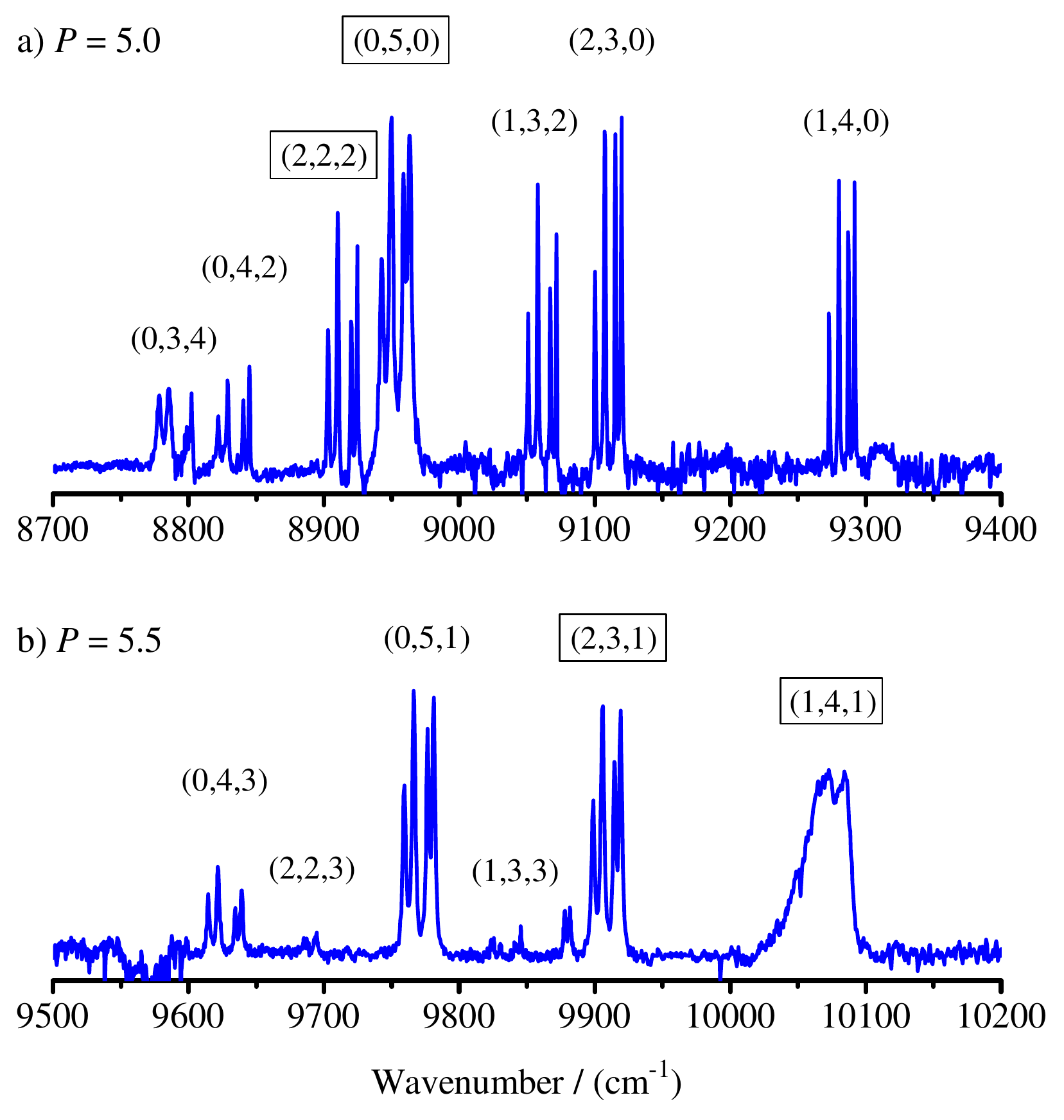}
        \caption{SEP scans over a) polyad 5.0 and b) polyad 5.5 with nominal assignments 
of the observed resonances 
        according to \lit{DCO_II_schinke}. 
        The four resonances selected for \ce{D} atom imaging are highlighted by frames 
around 
the state labels. 
With the pump laser tuned to the $\widetilde B \leftarrow \widetilde X,\ 0_0^0$, 
$^qR_0(0)$ line at $\nu_{\text{pump}}= \unit[38\,631.60]{cm^{-1}}$, 
state, 
the dump transition can reach four different rotational states, $N_{K_aK_c} = 0_{00},\ 
2_{02},\ 1_{10}$ and $2_{12}$, 
in each $\widetilde X$ vibrational state, explaining the four-line patterns observed in 
the SEP scans.
}
        \label{fig:sep}
\end{figure}

Four \ce{DCO} ($\widetilde X$) resonances were picked at random for 
investigating the KER spectra and \ce{CO} ($v, j$) product state 
distributions by the \ce{D} atom velocity map imaging experiment. 
Two states were taken from polyad $P = 5$ (at {wavenumbers} ${\tilde \nu_v}= 8\,902$ and 
$\unit[8\,942]{cm^{-1}}$) 
  and two from polyad $P = 5.5$ (${\tilde \nu_v} = 9\,896$ and $\unit[10\,065]{cm^{-1}}$). 
SEP spectra illustrating the observed vibrational states belonging to both polyads are 
shown in \autoref{fig:sep}. 

Following Keller \emph{et al.},\cite{DCO_II_schinke} the four resonances are nominally 
labeled 
as (0,4,2), (0,5,0), (2,3,1) and (1,4,1), in order of increasing energy for 
reference below.  
We emphasize, however, that these ``assignments'' have to be taken with caution. 
For example, the $\unit[8\,902]{cm^{-1}}$ resonance was reported as a mixture of 
\unit[48]{\%} (2,2,2) 
and \unit[32]{\%} (0,4,2), while the neighboring $\unit[8\,821]{cm^{-1}}$ resonance was 
found as a 
mixture \unit[41]{\%} (2,2,2) and \unit[32]{\%} (0,4,2). 
Further, the resonance at ${\tilde\nu_v} = \unit[10\,065]{cm^{-1}}$ nominally assigned as (1,4,1) 
is 
special because it shows pronounced interpolyad mixing with a highly dissociative 
resonance belonging to polyad $P = 6$, most likely (4,2,0).\cite{DCO_wei_2004}
{We note that we use the same notation $(v_1,v_2,v_3)$ for both zero-order 
states and resonance states. Which state is meant should either be clear from 
the context or the state is explicitly designated as resonance or zero-order 
state.}

\begin{figure}
        \includegraphics[width=\columnwidth]{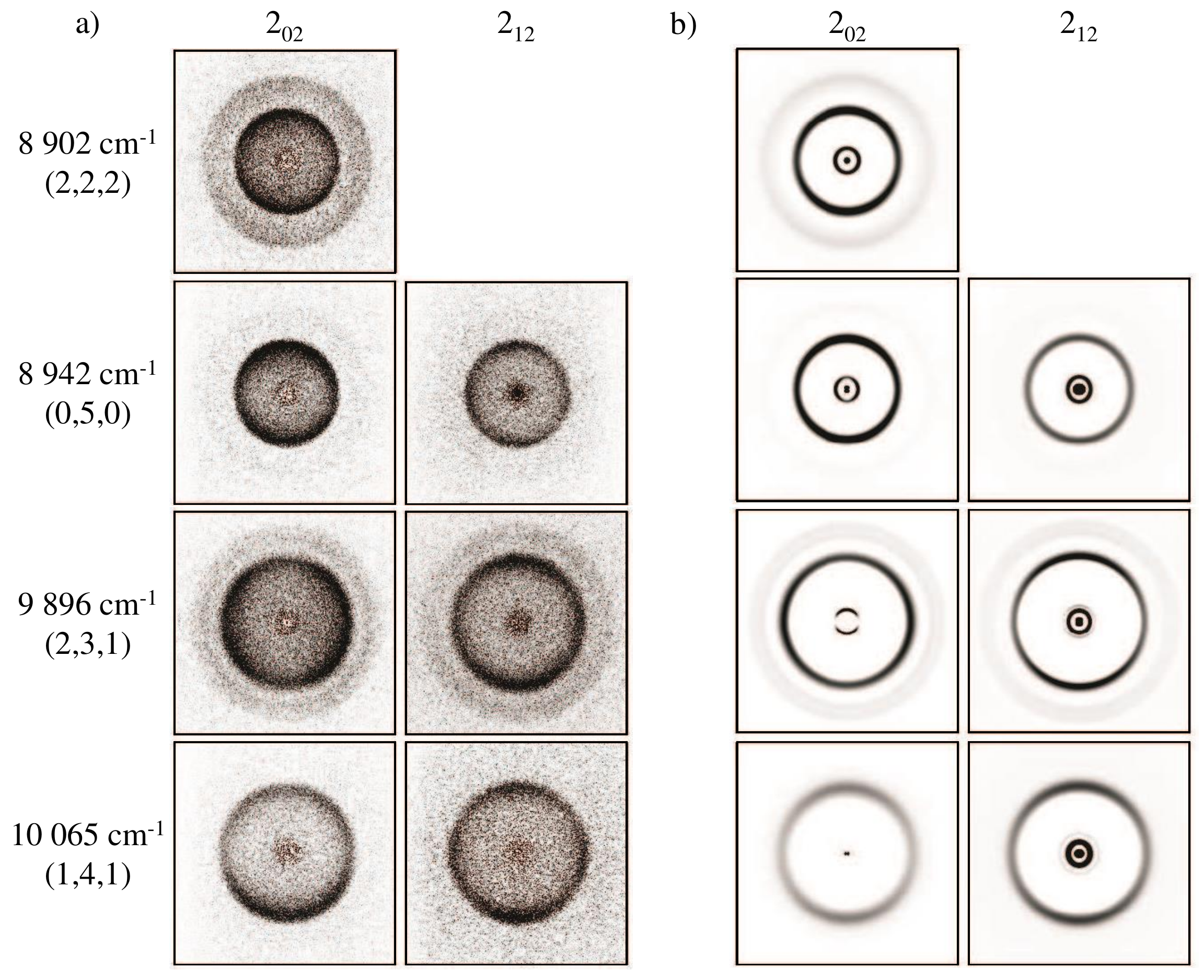}
        \caption{a) Measured two-dimensional (2D) \ce{D} atom velocity map images after 
excitation of the (0,4,2), (0,5,0), (2,3,1) and (1,4,1) DCO ($\widetilde X$) resonances 
in 
the 2$_{02}$ and 2$_{12}$ rotational states and b) reconstructed meridional slices 
through 
the respective three-dimensional (3D) \ce{D} atom recoil distributions. }
        \label{fig:images}
\end{figure}

The recorded D atom velocity mapped images after excitation of the 2$_{02}$ and 2$_{12}$ 
rotational states of the above four \ce{DCO} ($\widetilde X$) resonances are printed in 
\autoref{fig:images} a) together with their Abel inversions in \autoref{fig:images} b). 
The 
depicted meridional slices through the three-dimensional (3D) \ce{D} atom recoil 
distributions 
were obtained using the iterative regularization method with the projected Landweber 
algorithm.\cite{Renth2006} As can be seen, the raw images and the associated slices 
through the 3D recoil distributions differ substantially from vibrational resonance to 
resonance, while the results for the two rotational states are very similar, indicating 
the good reproducibilty of the data. For (0,5,0) and (1,4,1), the 3D slices show \ce{D} 
atoms  with comparably low recoil velocities. The allowed maximal available 
energies 
$E_{\text{avl}}$ is determined by the energy balance
\begin{equation}
  E_r +{ \unit{h} (\nu_{\text{pump}} - \nu_{\text{dump}})} = \Delta 
  {E}_0^0\text{(\ce{D-CO})} + 
E_{\text{avl}},
\end{equation}
where $E_r$ is the initial state's rotational state (here $E_r = 0$ for $N_{K_aK_c} = 
0_{00}$), 
$\Delta {E}_0^0\text{(\ce{D-CO})}$ is the asymptotic \ce{D-CO} dissociation 
energy,
and $\nu_{\text{pump}}$ and $\nu_{\text{dump}}$ are the {frequencies} supplied by the pump and 
dump laser pulses. Thus, 
dissociation of the (2,2,2) and (0,5,0) resonances may lead to \ce{CO} in $v = 0$ and $v 
= 1$, 
while the (2,3,1) and (1,4,1) resonances may also give \ce{CO} in $v = 2$. 
Evidently, however, (0,5,0) and (1,4,1) give mainly \ce{CO} in $v = 1$. Further, as 
by the narrow recoil widths, the \ce{CO}\,($v = 1$) acquires relatively low rotational 
excitation (quantum number $j$). 
In contrast, the images obtained from the (2,2,2) and (2,3,1) resonances show the 
formation of \ce{CO} in $v = 1$ and in $v = 0$, with broader rotational excitation in $v 
= 0$. 
The observed recoil anisotropies are negligible within experimental errors, consistent 
with resonance lifetimes ($\tau=\unit[0.8\,\text{--}\,5]{ps}$ inferred from the measured 
resonance 
widths) longer than the DCO rotational period.

Subsequent integration of the D atom images over the angular coordinates and 
transformation of the 
resulting recoil velocity distributions to the recoil translational energies with account 
of the fragment mass ratio and the appropriate Jacobian finally gave the total kinetic 
energy release (KER) spectra for the decaying resonances. 
The corresponding CO vibrational and rotational product state distributions follow by the 
energy balance.
The experimental KER spectra will be presented and compared with the 
present theoretical predictions in \autoref{sec:rovib_distr}.
The asymptotic \ce{D-CO} dissociation energy was assumed to have a value of 
$\Delta {E}_0^0 
= \unit[5\,450]{{h\, c\,}cm^{-1}}$.

\subsection{Simulation parameters}
\label{sec:res_params}
For the filter diagonalization, the initial wavefunction is taken from 
\lit{DCO_II_schinke} and takes the form of the following Gaussian (excluding 
normalization)
\begin{equation}
\begin{split}
 \Psi^{\text{FD}}&(R,r,\theta; t=0) \propto %
 \exp\{-[(R-R_0)/\alpha_R]^2\} \\
\times&\exp\{-[(r-r_0)/\alpha_r]^2\}\exp\{-[(\theta-\theta_0)/\alpha_\theta]^2\}
,
 \end{split}
\end{equation}
with the parameters $R_0 = \unit[3.05]{a_0}, r_0 = \unit[2.57]{a_0}, \theta_0 = 
138^\circ, \alpha_R = \unit[0.256]{a_0},  \alpha_r = \unit[0.195]{a_0},$ and 
$\alpha_\theta=7.19^\circ$. This initial wavepacket is propagated for $\unit[2]{ps}$.
The filter function is a Gaussian with duration of $\tau=\unit[1]{ps}$ (see 
\autoref{eq:filterfunction}), and the energy region of 
interest (see \autoref{sec:energies}) was divided into four regions, each with
a width of 
$\sim\unit[400]{{h\, c\,}cm^{-1}}$ and with typically $10$ energy-filtered basis 
functions. For each region, the number of 
basis functions was adapted to avoid an overcomplete basis.

The DVR basis set parameters are given in \autoref{tab:basis_parameters}. 
For the filter diagonalization, a smaller range in $R$ up to $R=\unit[10]{a_0}$ is used.
Due to a wrong 
asymptote of the SAG potential, a smaller 
coordinate range in $r$ was used for all simulations on that PES.
In coordinate $R$, we use the transmission-free complex absorbing potential 
{(CAP)} of 
\lit{cap_manolopoulos_2002} (taking the rational form, Eq.~(2.25) in that reference) with 
a
width of $\unit[5]{a_0}$ for the filter diagonalization and a width of $\unit[10]{a_0}$ 
for the decay dynamics. 
{This corresponds to starting positions of the CAP at $\unit[5]{a_0}$ and $\unit[18]{a_0}$, respectively. Similar values have been used previously for obtaining resonance states.\cite{DCO_II_schinke,HCO_mctdh_ndengue_2015,HCO_tremblay_2005} 
For selected states, we have done simulations with other CAP positions and widths and found no significant deviations for neither the resonance energies/widths nor the decay distributions.}
We use the following atomic masses: $\unit[12.0096]{Da}$ for \ce{C},\cite{iupac_amu}
$\unit[15.99977]{Da}$ for \ce{O} and $\unit[2.01410178]{Da}$ for \ce{D}.\cite{ame2016}

The wave-amplitude threshold used for the pruning of the filter diagonalization 
dynamics is $10^{-11}$, although a threshold of $10^{-8}$ would give the same results. 
This looser threshold is used for the decay dynamics. 
There, a {threshold} of $10^{-6}$ would have been sufficient.
The analysis line (\autoref{sec:theory_rovib}) is placed at $R_\infty = 
\unit[16]{a_0}$.
For all propagations, we employ a short iterative Arnoldi 
propagator\cite{sil_light_1986,mctdh_rev_meyer_2000} with an accuracy of
$10^{-10}$, in the form in which it is also 
implemented in the Heidelberg MCTDH package.\cite{mctdh_package}
For each resonance, the final propagation time and the norm of the wavefunction at that 
time can be found in Table i in the supplementary material.

We note that all parameters were chosen conservatively and are not well-optimized. 
Especially, the employed coordinate ranges are probably too large. They do not require 
an in-depth optimization because our DP algorithm does this automatically: If the 
wavepacket will not enter a certain region in coordinate space, no basis function 
will become active in that region during the propagation and computational 
costs will not increase, compared to a carefully optimized coordinate range. 
Further, this means that the CAP width needs not be tuned to increase computational 
efficiency (although, the stronger the CAP, the earlier the wavefunction is absorbed in 
the CAP region, and the fewer basis functions are needed). 
This gives the DP-DVR algorithm 
more of black-box character than conventional DVR dynamics. However, the other parameters 
still 
required convergence tests. 
The number of DVR functions, which determines the maximum momentum that can be described, 
can be easily determined by checking reduced densities of the wavefunction in momentum 
space.
For the filter diagonalization and decay dynamics of 
selected resonances, careful convergence tests with tighter parameters have been pursued 
and the stated final parameters, {including $R_\infty$}, were chosen 
based on conservative 
conclusions from these tests.%
\footnote{To avoid unnecessary computational costs, not all the resonances 
have been propagated with the final parameters. Further, the propagation for the filter 
diagonalization has been performed with a larger basis but the results with the basis 
from \autoref{tab:basis_parameters} are virtually identical.}

\begin{table}[!htbp]
\caption{Basis sizes $N$ and coordinate ranges for the three Jacobi coordinates $R, r$ 
and $\theta$
for the two employed PES. {The values in the ranges are given in units of 
  the Bohr radius $a_0$.\cite{conventions_quack}}}
\label{tab:basis_parameters}
\begin{ruledtabular}
 \begin{tabular}{lll|ll|ll}
      &\multicolumn{2}{c|}{$R$} & \multicolumn{2}{c|}{$r$} & 
\multicolumn{2}{c}{$\theta$}\\ \hline
 PES &$N$ & range & $N$ & range & $N$ & range\\
 \hline
 WKS\cite{HCO_I_schinke,HCO_II_schinke} & 210 & $[1.5,28]$ & 78 & $[1.5,6]$ & 60 & 
$[0,\pi]$\\
 SAG\cite{HCO_pes_song_2013} & -"- & -"- & 34 & $[1.5,3.49]$ & -"- & -"- \\
 \end{tabular}
 \end{ruledtabular}
\end{table}

\subsection{Energies and widths}
\label{sec:energies}
The computed resonance energies and widths on the WKS surface are given in 
\autoref{tab:energies} and compared with both 
{the experimental and theoretical} values from 
\lits{DCO_I_schinke,DCO_II_schinke}. 

Note that some energetically low-lying resonances (below ${\tilde\nu} =
\unit[8770]{cm^{-1}}$, ${\tilde\nu}$ is the vibrational resonance {wavenumber} relative to the 
ground state) 
in polyad $5$ and one in polyad $5.5$ 
are not included in our simulations. They were 
also not experimentally accessible. Here, we consider resonances in polyad $5$ and 
$5.5$ that are within a {range} of $\unit[8770]{cm^{-1}}$ and 
$\unit[10050]{cm^{-1}}$.

Compared to the computed {wavenumbers} from \lit{DCO_II_schinke}, ours are systematically 
smaller but agree within $\sim\unit[5]{cm^{-1}}$. The decay widths have a better 
agreement; there, except for 
resonance (0,3,4), the maximal absolute deviation is $\unit[1]{cm^{-1}}$. Note that 
Keller 
\textit{et al.} reported that their basis was probably too small to 
yield accurate calculations.\cite{DCO_II_schinke} Their basis used in the calculations 
for \ce{HCO}\cite{HCO_II_schinke} was significantly larger.
To our knowledge, there are no further calculations done for \ce{DCO} on the WKS surface.
However, theoreticians have used resonance calculations for \ce{HCO} on the WKS surface 
as a benchmark 
case.%
\cite{HCO_IV_schinke,HCO_poirier_2002a,HCO_mandelshtam_2002,HCO_tremblay_2005,
HCO_mctdh_ndengue_2015,HCO_pes_ndengue_2016} There, the deviations from the results of 
\lit{HCO_II_schinke} have a similar magnitude as the deviations shown here. 
To conclude, our DP-DVR method with filter diagonalization gives results that are within
reasonable agreement with the results from Keller \text{et al.} 
To allow for a better comparison, more extensive tests with \ce{HCO} should be done but 
this is not within the scope of this work.

\begin{table*}[!htbp]
    \caption{Comparison of computed resonance {transition wavenumbers $\tilde{\nu}$} (relative to 
the ground state) and widths $\Gamma$ from this work (DP) with the experimental 
(Exp.) and theoretical (WKS) results from \lits{DCO_I_schinke,DCO_II_schinke}.}
\label{tab:energies}
\begin{ruledtabular}
 \begin{tabular}{lc%
 rrrc|%
 rddl|%
 rclclclc %
 }
       &            &%
       \multicolumn{4}{c|}{$\tilde\nu/\unit{cm^{-1}}$} & %
        \multicolumn{4}{c|}{$\Gamma / \unit{cm^{-1}}$}\\\cline{3-4}\cline{5-7}\cline{8-10}
     P.\footnotemark[1] & label\footnotemark[2] &  %
     \multicolumn{1}{c}{Expt.\footnotemark[3]} & \multicolumn{1}{c}{WKS\footnotemark[2]} & 
\multicolumn{1}{c}{DP\footnotemark[4]}  & $\Delta$\footnotemark[5]& %
     \multicolumn{1}{c}{Expt.\footnotemark[3]} & \multicolumn{1}{c}{WKS\footnotemark[2]} & 
\multicolumn{1}{c}{DP\footnotemark[4]}  & $\Delta$\footnotemark[5]& %
     \multicolumn{8}{c}{decomposition of wavefunction\footnotemark[6]}\\\hline
5 & ((034)) & 8778  & 8780 & 8775 & 5 &  3.50  & 7.6 & 5.6 & 2 &38:& 034 & 30:& (411) & 16:& 222 & 16:& 124  \\
5 & ((042)) & 8821  & 8832 & 8830 & 2 & {<}2.00  & 1.1 & 1.1 & 0 &41:& 222 & 32:& 042 & 23:& 132  \\
5 & ((222)) & 8902  & 8901 & 8895 & 6 &  1.06  & 1.9 & 1.2 & 0.7 &48:& 222 & 32:& 042  \\
5 & (050)   & 8942  & 8953 & 8950 & 3 &  1.79  & 0.14 & 0.13 & 0.01 &44:& 050 & 33:& (140) & 15:& 230  \\
5 & (132)   & 9050  & 9031 & 9029 & 2 &  0.34  & 0.28 & 0.28 & 0 &59:& 132 & 24:& 042 & 11:& 222  \\
5 & (230)   & 9099  & 9099 & 9096 & 3 &  0.20  & 0.32 & 0.32 & 0 &57:& 230 & 32:& 050  \\
     5.5 & 027 & \multicolumn{1}{c}{---}  & 9235 & 9234 & 1 &  \multicolumn{1}{c}{---}  & 14 & 13 & 1 &100:& 027\\
5 & ((140)) & 9272  & 9251 & 9248 & 3 &  0.29  & 0.32 & 0.31 & 0.01 &64:& 140 & 20:& 230 & 14:& 050  \\
    5.5 & ((321)) & \multicolumn{1}{c}{---}  & 9496 & 9494 & 2 &  \multicolumn{1}{c}{---}  & 17 & 17 & 0 &57:& (321) & 32:& 035  \\
5.5 & (043) & 9614  & 9630 & 9629 & 1 &  2.30  & 1.5 & 1.4 & 0.1 &38:& (223) & 29:& 043 & 27:& 133  \\
5.5 & (223) & 9686  & 9690 & 9688 & 2 & {<}5.00  & 5.6 & 5.5 & 0.1 &52:& (223) & 25:& 043 & 13:& 035  \\
5.5 & ((051)) & 9757  & 9764 & 9762 & 2 &  0.83  & 0.66 & 0.64 & 0.02 &26:& 051 & 24:& 231 & 22:& 141 & 22:& 043  \\
5.5 & ((133)) & 9819  & 9807 & 9805 & 2 & {<}3.00  & 1.9 & 1.8 & 0.1 &45:& 133 & 21:& 043 & 21:& 051 & 11:& (223)  \\
5.5 & (231) & 9896  & 9893 & 9891 & 2 &  1.22  & 1.6 & 1.6 & 0 &56:& 231 & 29:& 051  \\
5.5 & ((141)) & 10065  & 10046 & 10044 & 2 &  6.00  & 3.9 & 3.9 & 0 &46:& 141 & 21:& 420 & 21:& 231  \\
 \end{tabular}
\footnotetext[1]{Polyad quantum number.}
\footnotetext[2]{Assignment according to \lit{DCO_II_schinke}. The more parentheses the 
assignment has, the more complicated the shape and the more difficult the assignment; see 
\lit{DCO_II_schinke} for details.}
\footnotetext[3]{According to \lit{DCO_I_schinke} and as given in \lit{DCO_II_schinke}.}
\footnotetext[4]{This work with the PES from \lits{HCO_I_schinke,HCO_II_schinke}.}
\footnotetext[5]{Difference between the theoretical results from \lit{DCO_II_schinke} and 
ours.}
\footnotetext[6]{Each entry: First part is contribution in percent, second part is 
zero-order state assignment; taken from \lit{DCO_II_schinke}.}
 \end{ruledtabular}
\end{table*}

In \autoref{tab:energies_SAG}, vibrational {wavenumbers} and widths for five selected 
resonances are 
shown for the SAG surface and compared against experiment and results using the WKS 
surface. For four of them, experimental KER spectra are available as well 
(\autoref{sec:res_exp}).
Compared to our WKS results, the {wavenumbers} differ from those 
obtained with the SAG surface by up to $\unit[54]{cm^{-1}}$. The widths differ by up to 
$\unit[3]{cm^{-1}}$. The WKS {wavenumbers} are closer to the experimental values and so are 
most of the widths --- except for resonance (0,5,0).
Note that some SAG states have components of the wavefunctions at linear geometries where 
Renner-Teller coupling and a conical intersection occur; see 
\autoref{sec:dp-dvr}. Using 
single-reference coupled-cluster calculations, the SAG surface was not optimized for 
linear geometries. %
Since both the {wavenumbers} and rovibrational distributions (\autoref{sec:rovib_distr}) from 
the WKS surface are typically closer to the experiment, we focus in the following on the 
WKS 
surface but mention the SAG results where they are available.\footnote{The PES of 
Ndengu{\'e} \textit{et al.} shows a better agreement to experimental data for many but 
not all resonances in \ce{HCO}. Some {wavenumbers} and widths are actually still better 
described by the WKS surface. Thus, it cannot be expected that this PES would give much  
improved results than the two PES studied here.}

\begin{table*}[!htbp]
    \caption{Comparison of computed resonance {transition wavenumbers $\tilde \nu$} and widths 
$\Gamma$ from this work with the PES from \lits{HCO_I_schinke,HCO_II_schinke} (WKS) and 
the PES from \lit{HCO_pes_song_2013} (SAG), together with the experimental (Exp.) results 
from \lits{DCO_I_schinke,DCO_II_schinke}.}
\label{tab:energies_SAG}
\begin{ruledtabular}
 \begin{tabular}{lc%
 rrrcc|%
 rdddd%
 }
       &            &%
        \multicolumn{5}{c|}{${\tilde\nu}/\unit{cm^{-1}}$} & %
        \multicolumn{5}{c}{$\Gamma / \unit{cm^{-1}}$}\\\cline{3-7}\cline{8-12}
P.\footnotemark[1] & Label\footnotemark[2] &  %
     \multicolumn{1}{c}{Expt.\footnotemark[3]} & 
\multicolumn{1}{c}{DP:WKS\footnotemark[4]} & \multicolumn{1}{c}{DP:SAG\footnotemark[5]}  & 
\multicolumn{1}{c}{$\Delta_\text{WKS}$\footnotemark[6]}& 
\multicolumn{1}{c}{$\Delta_\text{SAG}$\footnotemark[7]} &
     \multicolumn{1}{|c}{Expt.\footnotemark[3]} & %
\multicolumn{1}{c}{DP:WKS\footnotemark[4]} & \multicolumn{1}{c}{DP:SAG\footnotemark[5]}  & 
\multicolumn{1}{c}{$\Delta_\text{WKS}$\footnotemark[6]}& 
\multicolumn{1}{c}{$\Delta_\text{SAG}$\footnotemark[7]} \\ \hline
5 & ((042)) & 8821  & 8830 & 8849 & -9 & -28 &  {<}2.00  & 1.1 & 0.77 & {<}0.90 & {<}1.2  \\
5 & ((222)) & 8902  & 8895 & 8925 & 7 & -23 &   1.06  & 1.2 & 0.53 & -0.2 & 0.5  \\
5 & (050) & 8942  & 8950 & 8957 & -8 & -15 &   1.79  & 0.13 & 1.2 & 1.7 & 0.59  \\
5.5 & (231) & 9896  & 9891 & 9928 & 5 & -32 &   1.22  & 1.6 & 0.36 & -0.38 & 0.86  \\
5.5 & ((141)) & 10065  & 10044 & 10098 & 21 & -33 &   6.00  & 3.9 & 0.85 & 2.1 & 5.2  \\
 \end{tabular}
\footnotetext[1]{Polyad quantum number.}
\footnotetext[2]{According to \lit{DCO_II_schinke}.}
\footnotetext[3]{According to \lit{DCO_I_schinke}.}
\footnotetext[4]{This work with the PES from \lits{HCO_I_schinke,HCO_II_schinke}.}
\footnotetext[5]{This work with the PES from \lit{HCO_pes_song_2013}.}
\footnotetext[6]{Difference between Exp. and WKS results.}
\footnotetext[7]{Difference between Exp. and SAG results.}
 \end{ruledtabular}
\end{table*}

\subsection{Adiabatic rovibrational states}
\label{sec:adiabatic_repr}

To shed more light on the coupling of the different states, we look at the adiabatic 
potential energy curves in $R$ defined by the following eigenvalue problem in 
$\{r,\theta\}$:
\begin{equation}
\begin{split}
\left[-\frac{\hbar^2}{2\mu_{\ce{CO}} r^2} \partdd{}{r} + 
\left(\frac{1}{2\mu_{R} R^2} + \frac{1}{2\mu_r r^2}\right) \hat j^2 + 
V(R,r,\theta)\right] \\
\Ket{\chi_{n}(r,\theta; R)} %
= E_{n}(R) \Ket{\chi_{n}(r,\theta; R)},
 \end{split}
 \label{eq:ad_hamilt}
\end{equation}
where $\hat j^2$ is the angular momentum operator in $\theta$ and $V(R,r,\theta)$ is the 
WKS potential. $\mu_r$ is the reduced mass of \ce{CO}. A similar but 
one-dimensional Hamiltonian, averaged over $\theta$, was 
considered for \ce{HCO} in \lit{HCO_I_schinke}. 
For \ce{DCO}, the agreement found with this angle-averaged potential is less 
satisfactory.

\begin{figure*}%
 \includegraphics[width=.7\textwidth]{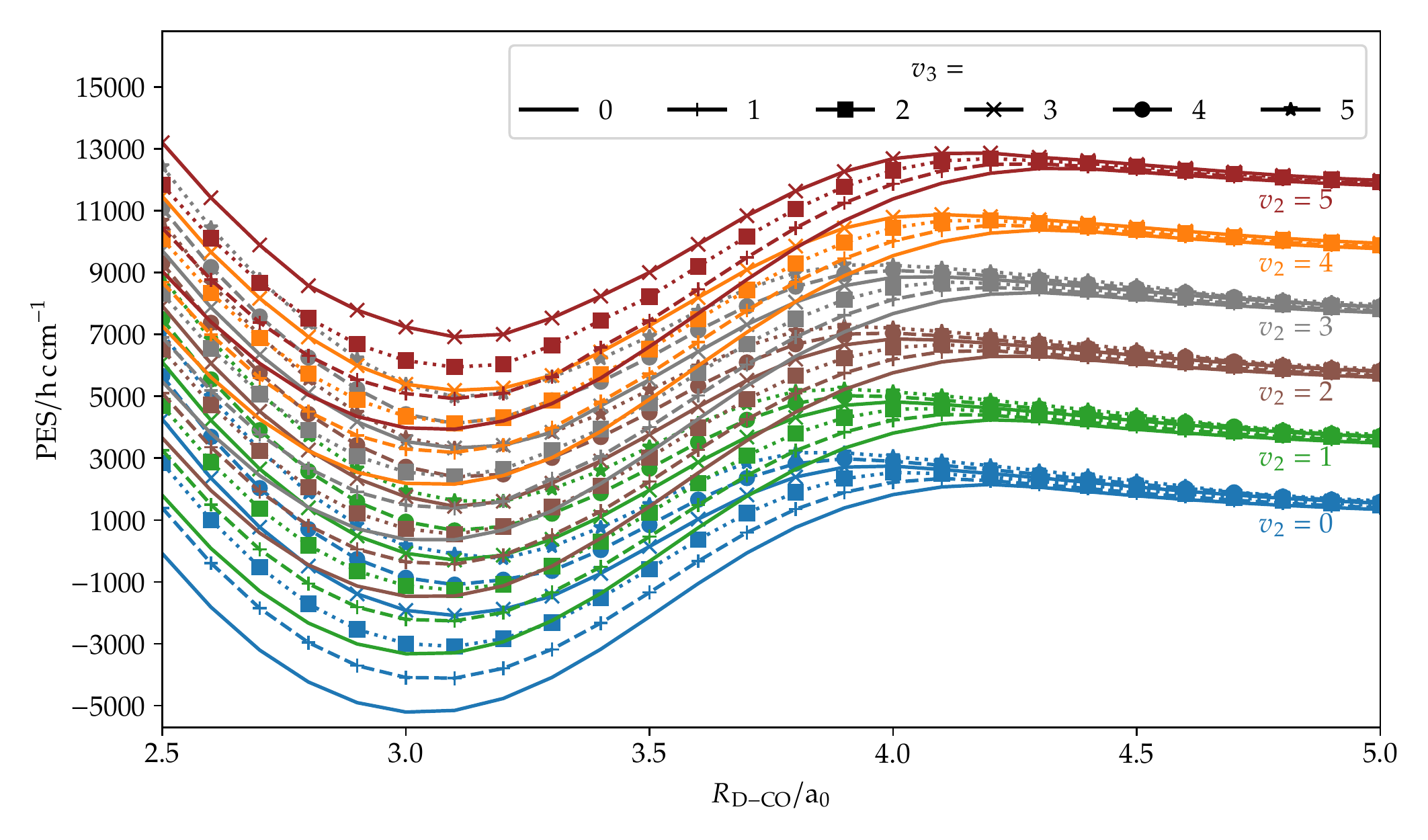}
 \caption{Adiabatic potential energy curves for the \ce{CO} vibrational problem as a 
function of
 \ce{D-CO} distance $R$ in Jacobi coordinates; see \autoref{eq:ad_hamilt}.
 The adiabatic curves are assigned \ce{DC=O} stretch $v_2$ and \ce{D-C=O} rotation 
$v_3$ quantum numbers 
 according to the character of the eigenstates.
 Note the occurrence of many avoided crossings where the character of the adiabatic 
states 
is ambiguous. 
 Further, note that only a selection of curves ($v_2\le 5$ and $v_3\le 5$) is depicted, 
 i.e., there are many more curves in this energy region. 
 For $v_2 \ge 4$, only curves for $v_3 \le 3$ are shown and they should be regarded 
as qualitative in the interaction region.
 }
 \label{fig:adiab_pes}
\end{figure*}

Selected adiabatic potential energy curves are shown in \autoref{fig:adiab_pes}. For 
$R < \unit[4]{a_0}$, there are many avoided crossings and as such strong couplings 
between states 
that have a different $v_2$ number.
For example, the curve for $v_2=3, v_3=0$ shows a coupling with $v_2=0, v_3=5$ and, at 
small $R$, even with $v_2=0, v_3=4$.
These curves can be used to qualitatively understand the mixing of the zero-order 
states: {Resonance state} (0,5,1) has contributions of (2,3,1), (1,4,1) and 
(0,4,3) (see 
\autoref{tab:energies}). 
The avoided crossing at $R\approx \unit[3.3]{a_0}$ for $v_2=5, 
v_3=1$ and $v_2=4, v_3=3$ explains the mixing of states (0,5,1) and (0,4,3).
Although the curve also overlaps with $v_2=3, v_3=5$, the mixing is less significant 
because the difference in the character of the two states is larger such that the 
coupling is reduced.
Due to the excitation in $v_1$, states (2,3,1) and (1,4,1) mix with (0,5,1) as well.

It would strongly simplify the analysis of the decomposition of the states if the 
adiabatic states $\Ket{\chi_{n}(r,\theta; R)}$ were (quasi-)diabatized such that the 
character of the states does not change for all considered values of $R$. 
A projection of the resonance states onto the diabatized states would give a quantitative 
decomposition in terms of quantum numbers $v_2$ and $v_3$. However, we were not able to 
obtain useful diabatic states that have a conserved nodal pattern for all values of 
$R$; see Section iii in the supplementary material for more details.

\subsection{Product distributions}
\label{sec:rovib_distr}
The experimental and convoluted theoretical KER spectra 
for resonance {states} (2,2,2), (0,5,0), (2,3,1) and (1,4,1) are shown in 
\autoref{fig:KER}.
All theoretical KER spectra without convolution are shown in the supplementary 
information (Figures iv -- x{i}v).
The agreement between the theoretical and the experimental spectra is good.
For (0,5,0) the spectrum from the SAG surface agrees better with experiment.
Whereas the spectrum from the WKS surface shows a bimodal distribution,
both experimental and SAG spectra show a monomodal rotational distribution for $v=0$.
This is in accord with the better agreement of the resonance width.  For 
the WKS surface, the width is too small; compare with \autoref{tab:energies_SAG}.
However, for the other three resonances, the WKS surface shows better agreement with the 
experiment, both in terms of peak positions and intensities. Compared to the experimental 
results, some, but not all, peaks exhibit a minor shift. This cannot be explained simply 
by a different asymptotic dissociation energy because there is no consistent trend.
Instead, the shifts probably arise because the KER spectra are very sensitive to the 
shape of the PES, in particular near the transition region.

\begin{figure*}
 \includegraphics[width=.7\textwidth]{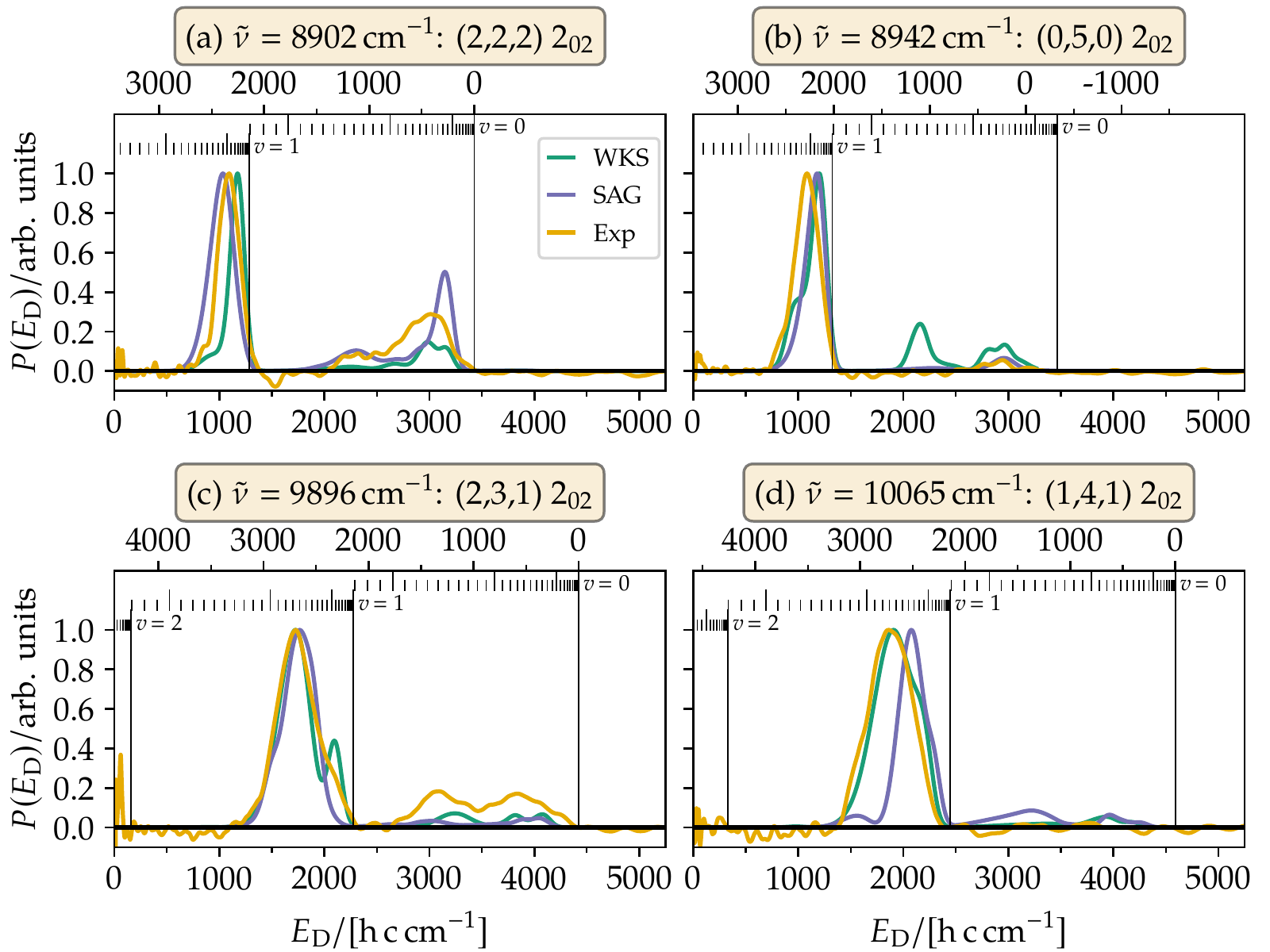}
 \caption{
Experimental (Exp) and convoluted theoretical (WKS and SAG) total kinetic energy release 
spectra $P(E_{\ce{D}})$ and corresponding \ce{CO} rovibrational product state 
distributions 
for the experimentally investigated \ce{DCO} ($\widetilde X$) resonance states.
WKS\cite{HCO_I_schinke,HCO_II_schinke} and SAG\cite{HCO_pes_song_2013} stands for the two 
employed PES.
The vertical lines indicate the maximal available energies $E_{\text{avl}}$ for 
formation of \ce{CO} in $v$ = 0, 1 and 2 and the energies of the \ce{CO} product 
rotational ($j$) states.
The \ce{CO} rovibrational energies corresponding to the vertical lines are obtained from 
computations with the WKS PES.
The theoretical curves are convoluted with a Gaussian, $\exp[-(E/\sigma)^2/2]$, 
with an arbitrarily chosen width  of $\sigma=\unit[60]{{h\, c\,}cm^{-1}}$.
}
 \label{fig:KER}
\end{figure*}

To allow for a easier comparison of the theoretical results, we have converted the 
energetically resolved 
spectra to distributions resolved rovibrationally by the asymptotic \ce{CO} 
fragment (see \autoref{sec:theory_rovib}). 
Two of the distributions (for resonances (2,3,1) and (0,5,0)) are shown in 
\autoref{fig:bars} 
(black/darker 
bars).
The rest of the distributions is shown in the supplementary material (Figures xv -- 
xx{i}v).
Except for (0,2,7), all computed resonances have major contributions for $v=1$ (compared 
to $v=0$) and 
negligible contributions for $v=2$. As in \ce{HCO},\cite{HCO_I_schinke,dixon_1992} 
the distributions typically are multimodal. Especially the resonances with large initial 
quantum 
numbers in $v_2$ and $v_3$ show a complicated multimodal pattern for a \ce{CO} 
stretch quantum number of $v=0$. 
The multimodal pattern is in agreement with semiclassical 
estimates used for \ce{H2O} and 
\ce{HCO}.\cite{H2O_schinke_1991,HCO_I_schinke,HCO_III_schinke}

To evaluate the changes of these distributions during the decay, we projected the part of 
the initial ($t=0$) resonance that lies in the dissociation region (we chose 
$R\ge\unit[5]{a_0}$) 
onto the asymptotic rovibrational states; see \autoref{eq:P_Ed_analysis_line_vj}. 
Characteristic examples of such
distribution changes are shown as gray/brighter bars 
in \autoref{fig:bars}. 
The relative amount of the initial wavepacket that lies in this asymptotic region
ranges from $7\%$ (resonance (0,5,0)) to $33\%$ (resonance (3,2,1)). For most resonances, 
this 
ratio lies between  $13\%$ and $20\%$.
In most cases, the qualitative features of these asymptotic distributions 
do not change during the dynamics, hence the projection at $t=0$ provides a very 
reasonable estimate of the 
rovibrational product distribution, even though the major part (typically more than 
$80\%$) of the initial wavepacket resides in the interaction region.
Some resonances exhibit a shift to higher (lower) $j$ values for $v=0$ 
($v=1$).

Only for those resonances that have a small decay width ((0,5,0), (1,3,2), (2,3,0) and 
(1,4,0); (2,3,1) on 
the SAG surface), we see larger deviations, see, e.g., resonance (0,5,0) in 
\autoref{fig:bars}. A small decay width means a {long} decay time and as 
such more 
time for interference processes etc.~which lead to these more 
significant asymptotic pattern changes. 
This follows from an investigation of the non-adiabatic coupling elements of the 
adiabatic, $R$-dependent 
rovibrational states (see \autoref{sec:adiabatic_repr}). It shows that the different 
states couple with each other even for $R>\unit[4.5]{a_0}$, especially those with 
the same quantum number $v_2$; compare with \autoref{fig:adiab_pes} and Section 
iii in the supplementary material.
Further, the stated resonances are those where less than $10\%$ of the
initial wavepacket lies in the interaction region (naturally, a small decay width
causes a smaller fraction of that resonance in the asymptotic region). 

\begin{figure}
\includegraphics[width=\columnwidth]{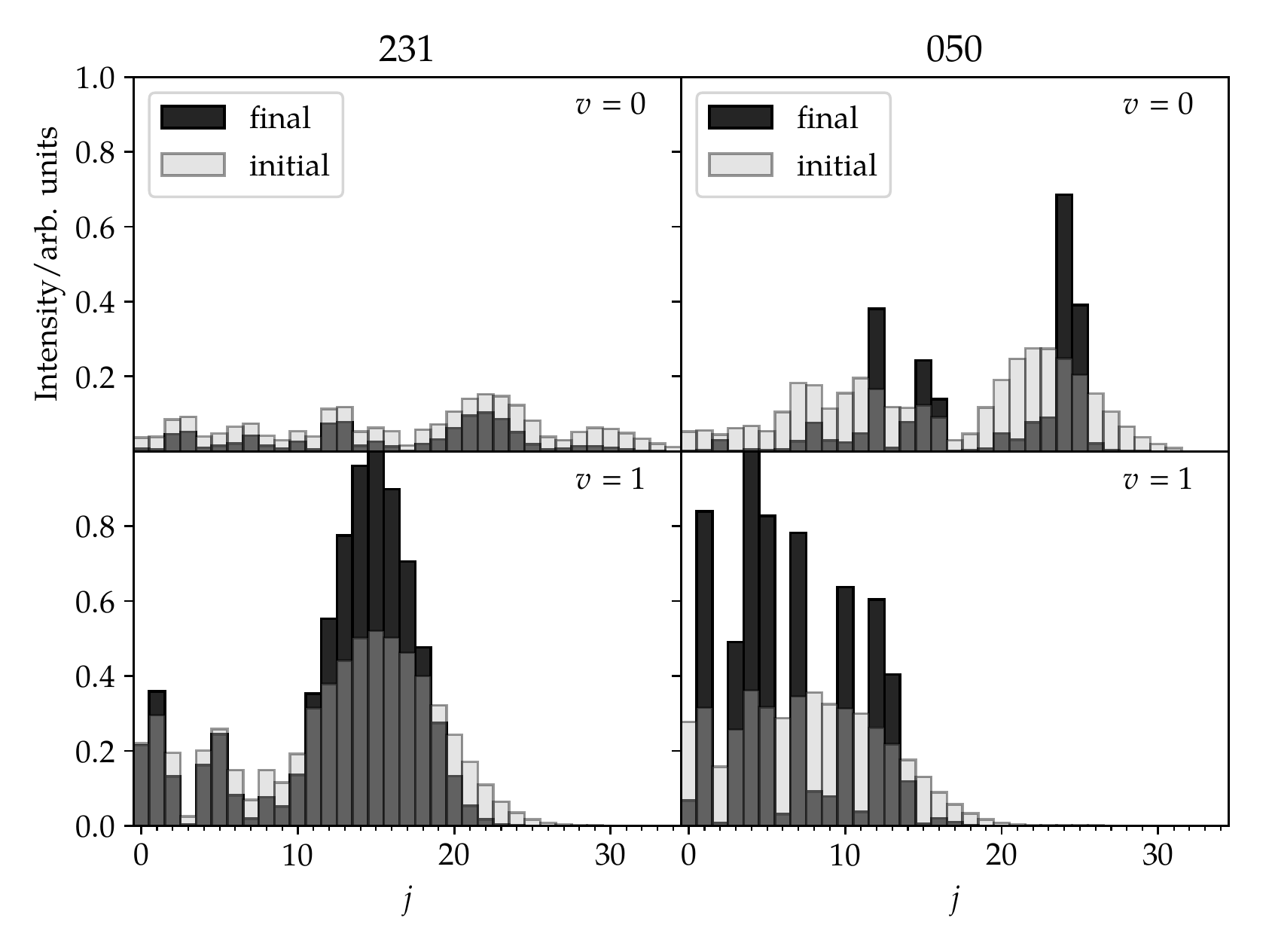}
\caption{Asymptotic distributions resolved by \ce{CO} vibration-rotation
  quantum numbers, for resonances (2,3,1) (left) and (0,5,0) (right) using the WKS 
PES.\cite{HCO_I_schinke,HCO_II_schinke}
The upper (lower) panels show the distributions for stretch quantum number $v=0$ 
($v=1$) versus 
different rotational quantum numbers $j$ of the asymptotic \ce{CO} fragment (bend 
quantum numbers for the undissociated state).
The distributions are scaled such that the maximal value is $1$.
The darker bars show the distributions for the final wavepacket, 
$\ket{\Psi(t\to\infty)}$,
the brighter bars for the initial wavepacket, 
$\ket{\Psi(t=0)}$, both in the asymptotic region.
}
\label{fig:bars}
\end{figure}

\subsection{Decay processes}
\label{sec:decay_processes}

We now analyze and discuss mechanisms of resonance decay, i.e.,
the occurrence of IVR during the decay.
{For that, the coordinate range in $R$ is increased (compare with 
  \autoref{sec:res_params}) such that the resonance states obtained from filter 
diagonalization are no longer eigenstates and will spread in $R$ and decay.}
{Note that we analyze here the part of the wavepacket that remains in the 
interaction region during the decay. A loss in a 
quantum number during the IVR means that the corresponding energy is moved from 
the interaction region to the continuum.
}

Using a polyad model Hamiltonian, the Temps group has already done an IVR
analysis.\cite{DCO_polyad_troellsch_2001,DCO_polyad_renth_2003}
During dissociation, the bending motion (quantum number 
$v_3$) turns into rotational motion of the \ce{CO} fragment (quantum number $j$). 
Semiclassically, 
four quanta 
need to be transferred to the \ce{D-C} stretch degree of freedom. The \ce{C-O} 
stretch quantum number $v_2$ remains (approximately) conserved and turns into $v$.

Here, we consider the \textit{ab initio} Hamiltonian, where an in-depth and quantitative
analysis is not possible.
As already mentioned in \autoref{sec:intro}, the assignment in terms of 
zero-order 
eigenstates and their vibrational quantum numbers is problematic. This was already 
visible in \autoref{tab:energies} above, and it becomes even more obvious when
analyzing the time-dependent wavefunctions further: Already counting the
number of nodes (by plotting $\Re{(\Psi)}$) or counting  the number of pronounced lobes 
(by plotting $|\Psi|^2$) can lead to different assignments because the intensities of the 
lobes vary significantly.
In this sense, any assignment should be regarded as approximate, and 
the IVR mechanism should be regarded as qualitative.

In the following, the assignment is performed based on counting the number of lobes. 
Qualitatively, we reproduced the assignment from \lit{DCO_II_schinke} (see 
\autoref{tab:energies}). 
The assignment was done based on cuts of Jacobi coordinates that have been 
shifted and rotated such that the first excited adiabatic state retained 
its orientation for different cut values (see also \autoref{sec:adiabatic_repr}).
Further, we plotted cuts in Eckart bond coordinates, i.e., coordinates $\{X, Y, 
\phi\}$, where $X$ ($Y$) is the \ce{C-D} (\ce{C-O}) distance and $\phi$ the angle 
$\sphericalangle(\ce{DCO})$ between the corresponding vectors $\vec{X}$ and $\vec{Y}$.
Here, we solely show cuts of the wavefunction in Eckart coordinates, since
they provide clearer nodal patterns in the decisive interaction region.
Throughout, all cuts are performed along coordinate 
values that show significant contributions to the wavefunction.

Note that we analyze the wavefunction in the interaction region, that is, we analyzed 
that  part of the wavefunction that remains {bound} during the studied 
propagation 
times. Hence, the following analysis is not directly comparable with the rovibrational 
product state distributions shown in \autoref{sec:rovib_distr}.
The direct conversion of the bound to the unbound part in the continuum 
is hard to analyze since it happens over a wide coordinate range and over long 
propagation times and since the initial wavefunctions already extends into the 
dissociation region. Indeed, already the initial wavefunctions qualitatively contains all 
required components in the asymptotic region (see \autoref{sec:rovib_distr}), 
although these asymptotic parts typically contribute only 13--20\% to the
total resonance wavefunction.

\subsubsection{Polyad 5}
\label{sec:mech_5}

\begin{figure}
 \includegraphics[width=\columnwidth]{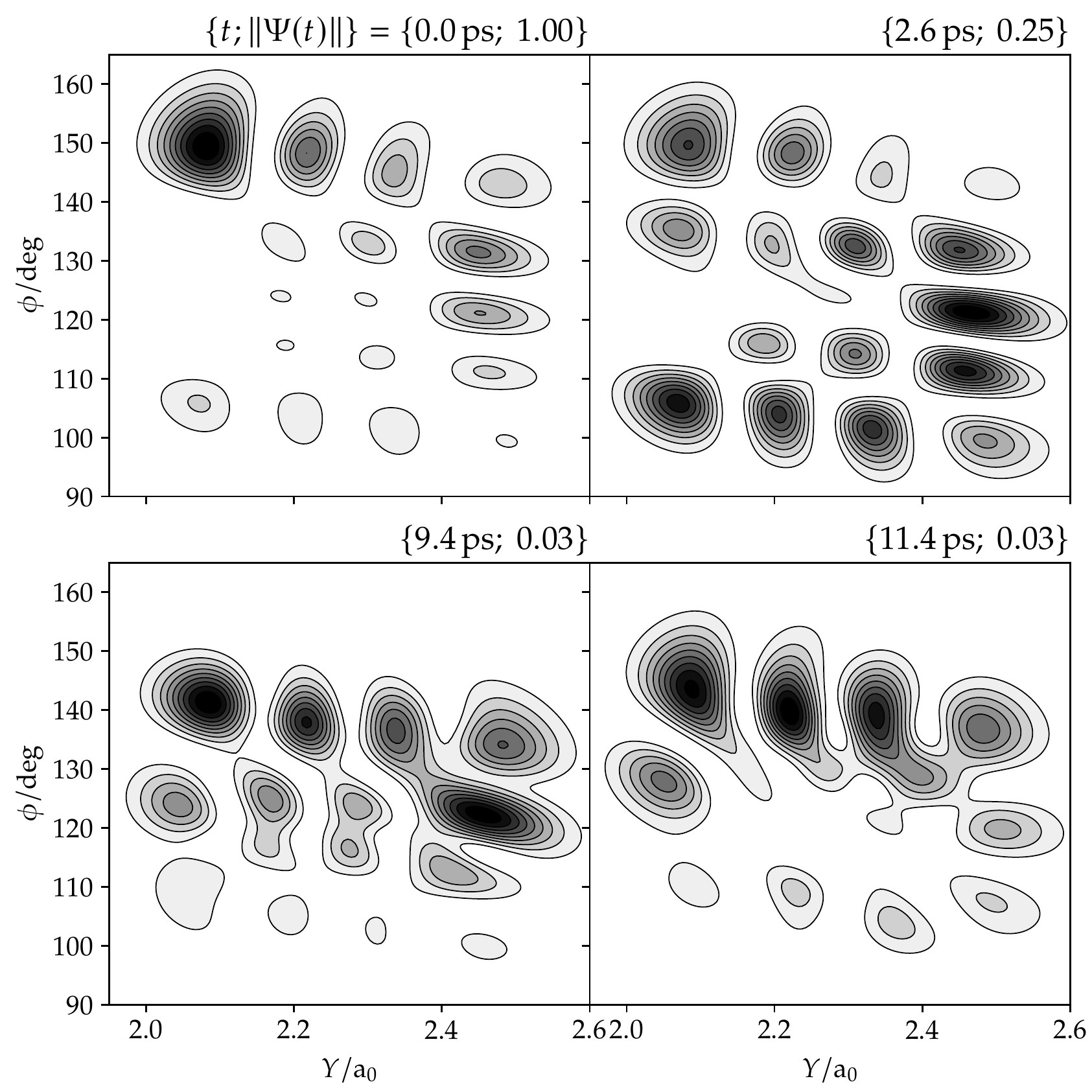}
 \caption{Contour values of $|\Psi(X,Y,\phi; t)|^2$ at $X=\unit[2.2]{a_0}$ 
and different times, for the (0,3,4) resonance. $X$ is the \ce{D-C} distance, $Y$ 
the \ce{C\bond{3}O} distance and $\phi$ the \ce{D-C=O} bending angle in Eckart bond
coordinates.
 For each panel, the shown function is normalized to have the maximal value of $1$. For 
a comparison of the relative intensity, see the norm values in the panel captions. 
 }
 \label{fig:decay_034}
\end{figure}

\begin{figure}
 \includegraphics[width=\columnwidth]{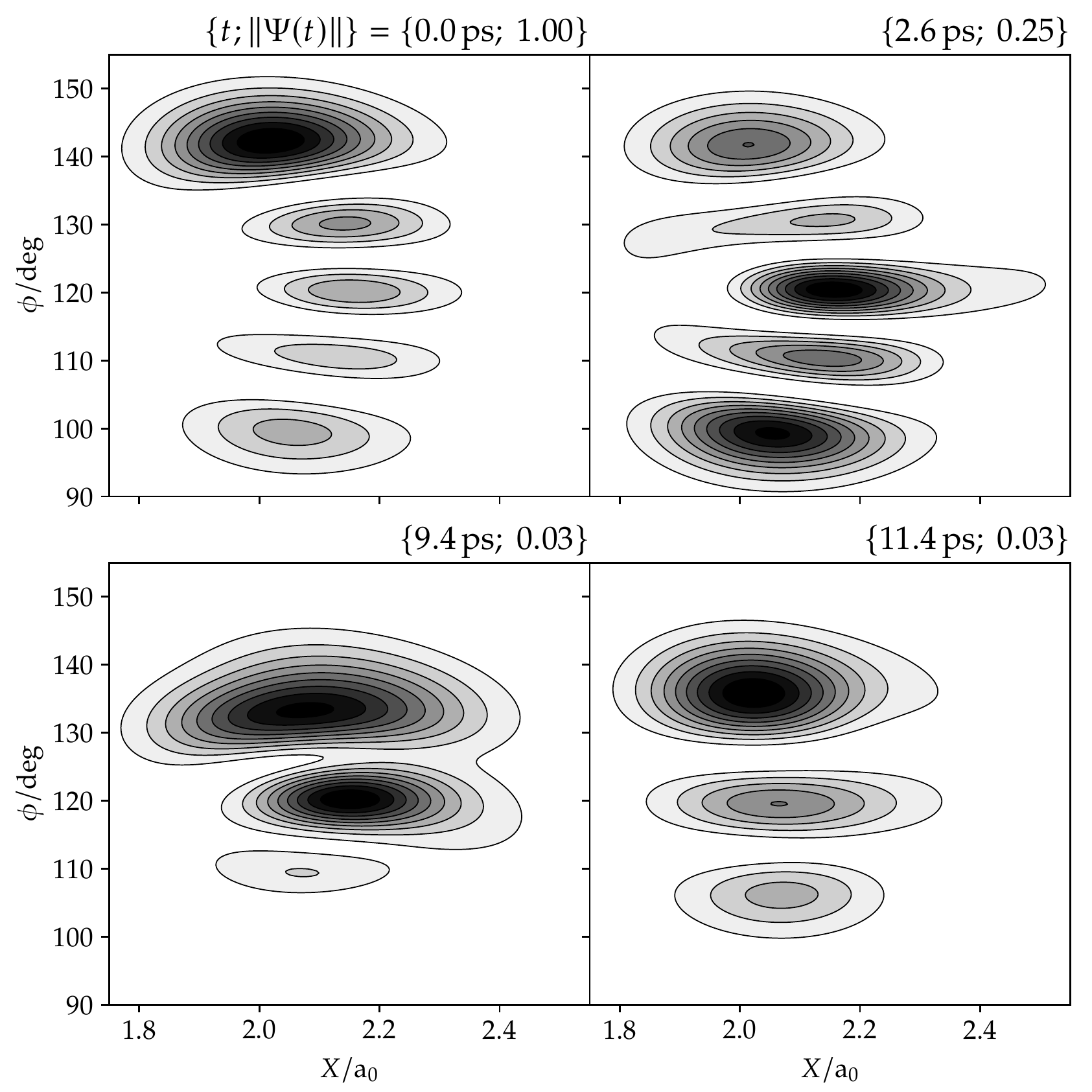}
 \caption{Same as \autoref{fig:decay_034} but showing cuts of $|\Psi(X,Y,\phi; t)|^2$ 
at $Y=\unit[2.53]{a_0}$.
$X$ is the \ce{D-C} distance, $Y$ the \ce{C\bond{3}O} distance and $\phi$ the \ce{D-C=O} 
bending 
angle in Eckart bond coordinates.
}
 \label{fig:decay_034_R}
\end{figure}

Cuts of the decay of the (0,3,4) resonance in the plane of \ce{C-O} stretching 
(\ce{D-C} stretching) and \ce{D-C-O} bending are shown in Eckart bond coordinates in 
\autoref{fig:decay_034} (\autoref{fig:decay_034_R}). 
The upper left panel of \autoref{fig:decay_034} shows a part of the initial 
resonance that can be assigned as $v_1 = 3$ and $v_3=4$ (there are four lobes in the 
\ce{C-O} distance $Y$ and  five lobes in the bending angle $\phi$). 
Here, the assignment is already ambiguous, because there is no contiguous nodal pattern. 
However, an inspection of different cuts and in different coordinates (see 
\autoref{sec:decay_processes}) verifies the assignment. 
The upper left panel of \autoref{fig:decay_034_R} shows a part of the initial 
resonance that can clearly be assigned as $v_{{1}}=0$ and $v_3=4$.  
Note that the other zero-order components are typically visible in different coordinate 
ranges such that there are regions where the different components do not significantly 
overlap.

A distinctive change in the vibrational structure happens only after the norm 
of the wavefunction has decreased to $0.15$ or less. 
{A decrease of the norm means a decay of the wavefunction because it 
  entered the region of the CAP.}
This is also the case for 
the other resonances studied.
Before IVR takes place, there are oscillations between the different zero-order components 
of the wavefunction, that is, the intensities of the lobes 
change.
This can be seen in the plots shown in the upper two panels of
\autoref{fig:decay_034}, although the oscillations are typically less intensive.

At propagation times when the norm is below $0.15$,
a decrease of the bending quantum number $v_3$ from $4$ to $2$ happens; see 
\autoref{fig:decay_034} and \ref{fig:decay_034_R}.
Neither an increase in $v_1$ nor a change in $v_2$ is visible in the interaction region.
This is valid also for other cuts of the wavefunction where other 
zero-order states show contributions. For the zero-order $411$ contribution to the 034 
resonance,
there is an additional decrease of $v_1$ to $2$.
The oscillations of the different zero-order states, the approximate 
conversion of quantum number $v_2$ and the decay of $v_3$ are all in agreement with 
the analysis done by a polyadic model 
Hamiltonian.\cite{DCO_polyad_troellsch_2001,DCO_polyad_renth_2003}

For the resonances (0,4,2), (2,2,2), (0,5,0), (1,3,2), (2,3,0), (1,4,0) (polyad 5) and 
(0,4,3) (polyad 5.5), 
only minor oscillations of the lobes occur.%
\footnote{
  For (0,4,2), the resonance {state} computed with the SAG PES does show an 
IVR with a 
decrease in $v_3$ to $2$ and $1$, but the initial state has a significant 
contribution near the linear configuration of \ce{DCO}. 
Since linear configurations are not properly described (see text), this result has to be 
taken with caution.
}
No IVR is visible and the 
oscillations are much less pronounced than that shown in the upper panels of 
\autoref{fig:decay_034} for (0,3,4).
To some extent, this can be explained by the small decay rate of the 
resonances. Except for (0,4,3) (polyad 5.5), they all have a decay width of $\Gamma \le 
\unit[0.32]{cm^{-1}}$. In polyad 5.5, resonance (0,4,3) has the second smallest value of 
$\Gamma$. A small reaction rate is correlated with a hindered 
IVR.\cite{DCO_polyad_renth_2003} However, resonance (0,5,1), the resonance with the 
smallest 
value 
of $\Gamma$ ($\unit[0.64]{cm^{-1}}$) in polyad 5.5, does show a decay 
mechanism  (see \autoref{sec:mech_55}) but also lies higher in energy and has an evenly 
spread-out distribution of four zero-order components such that the coupling to other 
states may be larger; see also below.

\begin{figure}
 \includegraphics[width=\columnwidth]{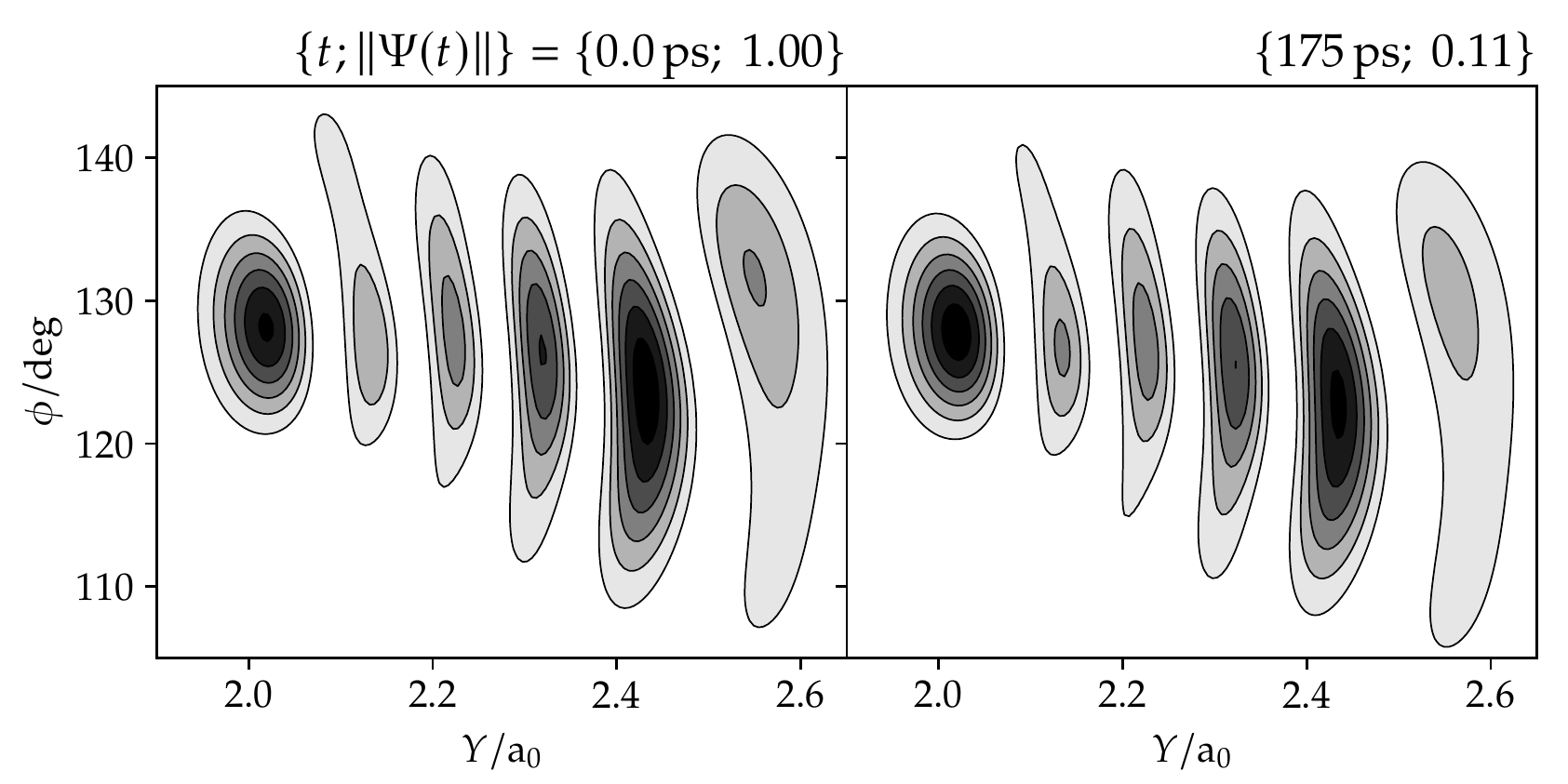}
 \caption{Same as \autoref{fig:decay_034} but showing cuts of $|\Psi(X,Y,\phi; t)|^2$ 
at $X=\unit[2.15]{a_0}$ for the (0,5,0) resonance.}
 \label{fig:decay_050}
\end{figure}

To show how insignificant the temporal change in wavefunction structure can be, consider 
the 
cut of the (0,5,0) resonance shown in \autoref{fig:decay_050}.
Even after propagating for \unit[175]{ps} and after decay of the norm 
to a value of $0.11$, the structure does not change.
For checking our results, we have propagated the (0,5,0) state using smaller coordinate 
ranges 
and 
a higher propagator accuracy with a standard DVR code (without DP) until \unit[300]{ps}, 
where the norm has dropped down to a value of $0.02$. Even then, no significant change in 
the 
structure of the wavefunction is visible. Additionally, 
our DP-DVR results are confirmed by this conventional DVR run. 

For a polyadic model Hamiltonian, Temps \textit{et al.}~studied the decay 
of a \emph{pure} (0,5,0) state.\cite{DCO_polyad_renth_2003} The state turned, among 
others, into (1,4,0) and (2,3,0) within less than $\unit[0.5]{ps}$.
Here, already the \emph{initial} resonance state consists of these components 
(not seen in the cuts shown in \autoref{fig:decay_050}); see \autoref{tab:energies}. 
Hence, the \emph{initial} state already has all main zero-order 
components needed for the IVR to happen. This explains the small 
oscillations in the nodal pattern.

This further means that the zero-order states included in the initial 
resonance states are strongly coupled whereas they couple weakly to other 
states.
The strong coupling of the states within one polyad is stressed in \lit{DCO_I_schinke}.
Indeed, resonances (0,4,2), (2,2,2) and (1,3,2) consist of the corresponding zero-order 
states and so do resonances (0,5,0), (1,4,0) and (2,3,0); see \autoref{tab:energies}. 
In contrast, resonance (0,3,4) shows an IVR. However, its decay rate is much larger and 
the resonance has components of (2,2,2) but also of the different zero-order states 
(4,1,1) and (1,2,4). 
The strong coupling between states $\{(0,4,2), (2,2,2), (1,3,2)\}$ and $\{(0,5,0), 
(1,5,0), (2,3,0)\}$ is in 
agreement with the parameters from the polyadic model 
Hamiltonian.\cite{DCO_polyad_troellsch_2001}

Considering the insignificant changes of the wavefunction within the interaction region, 
it 
may seem counterintuitive that for a subset of these states the asymptotic
rovibrational distributions of the \ce{CO} fragment do change significantly in
time. However, as pointed out in \autoref{sec:rovib_distr},
the rovibrational states still couple outside the typical interaction region where $R< 
\unit[4.5]{a_0}$, such that the dynamics in the asymptote and in the interaction 
regions are, to first order, not directly related. In fact, from
\autoref{fig:adiab_pes}, it may be argued that the distribution of the
adiabatic states on the energy axis, and thus also their couplings, strongly
change between $R<\unit[4.0]{a_0}$ and $R>\unit[4.0]{a_0}$, supporting
different dynamical behaviors.

As a side remark: On the SAG surface, resonance (0,5,0) does show an IVR with an 
increase in $v_3$ to $1$ (with intermediate numbers up to $v_3=3$) and a 
decrease in $v_2$ up to $v_2=2$.
However, the character of this resonance state also differs 
from the state  computed on the WKS surface. Most importantly, the initial state already 
has some  contributions of $v_3=1$ which the state on the WKS surface does not have (see 
\autoref{tab:energies}).
Note that, in this case, $\Gamma$ is larger and, like the KER, closer to the experimental 
value on the 
SAG surface. Resonances with excitations in both $v_2$ and $v_3$ exhibit larger decay 
rates than those with excitations only in $v_3$.\cite{DCO_II_schinke}

\subsubsection{Polyad 5.5}
\label{sec:mech_55}

In polyad 5.5, resonance (0,2,7) decays rapidly with a decrease in $v_2$. However, it has 
a 
strong contribution at the linear geometry. Due to the neglect of Renner-Teller coupling 
in this simulation and since resonance {states} with high bending motion 
are experimentally hard 
to 
measure, we have not analyzed this state further in the present study.

As already mentioned in \autoref{sec:mech_5}, resonance (0,4,3) shows only minor 
oscillations 
between the zero-order components of the initial state.
Resonance (2,2,3) shows a decrease in $v_3$ from $3$ to $\{1,0\}$ (both $1$ and $0$)
and an \emph{increase} in $v_2$ from $4$ to $5$.
However, due to a component of $v_3=5$, it shows also contributions at linear geometry 
and the results have to be taken with caution.
As for the previous resonances 
where a clear IVR takes place, (0,5,1) shows a decrease in 
$v_3$
whereas $v_2$ remains mostly conserved. %
Likewise, 133 decays to $v_3=\{2,1\}$ 
and the zero-order (0,5,1) state initially 
contributing by $\sim~21\%$ (\autoref{tab:energies}) gets more populated.

\begin{figure}
 \includegraphics[width=\columnwidth]{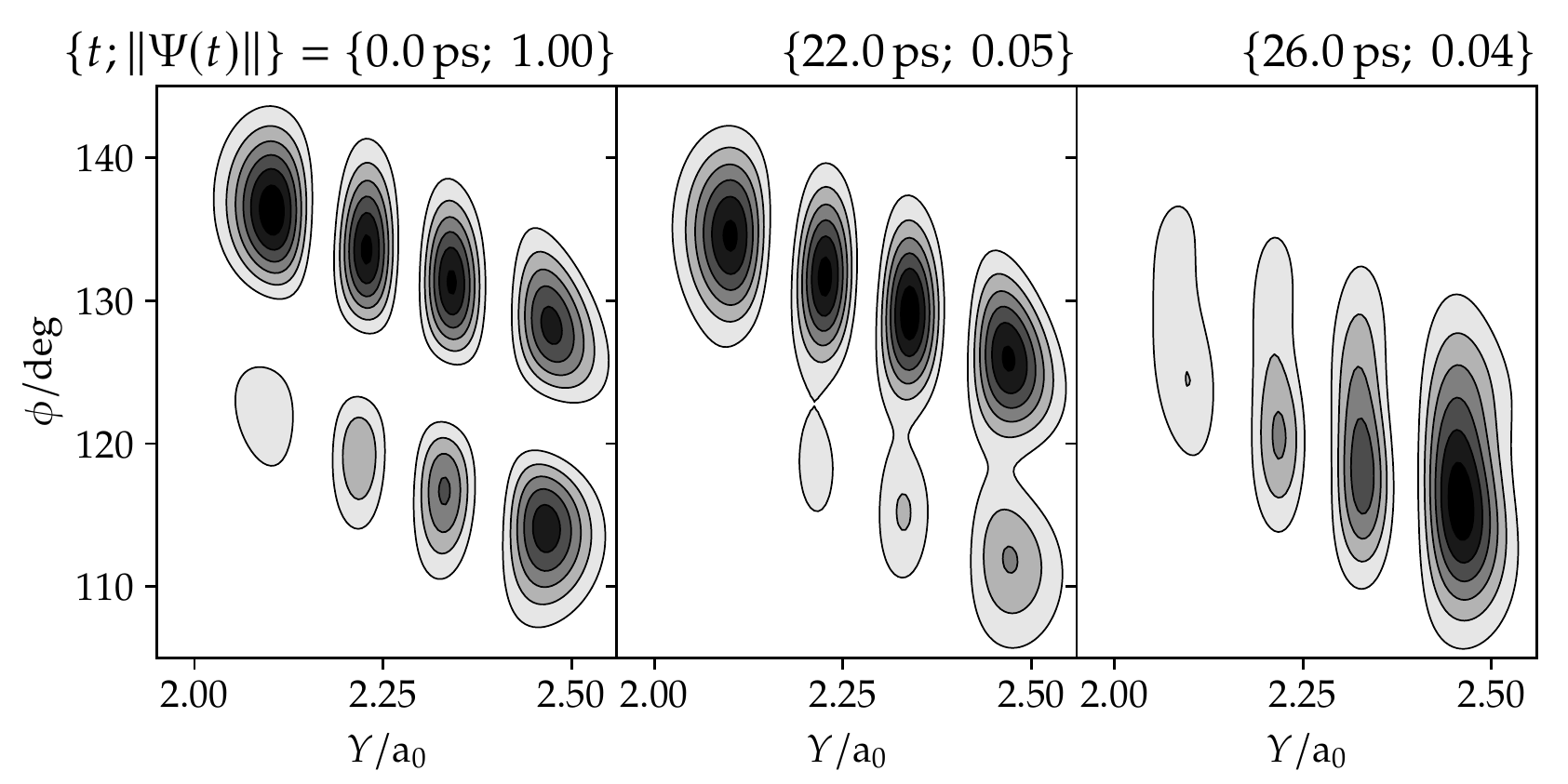}
 \caption{Same as \autoref{fig:decay_034} but showing cuts of $|\Psi(X,Y,\phi; t)|^2$ 
at $X=\unit[1.85]{a_0}$ for the (2,3,1) resonance.}
 \label{fig:decay_231}
\end{figure}

\begin{figure}
 \includegraphics[width=\columnwidth]{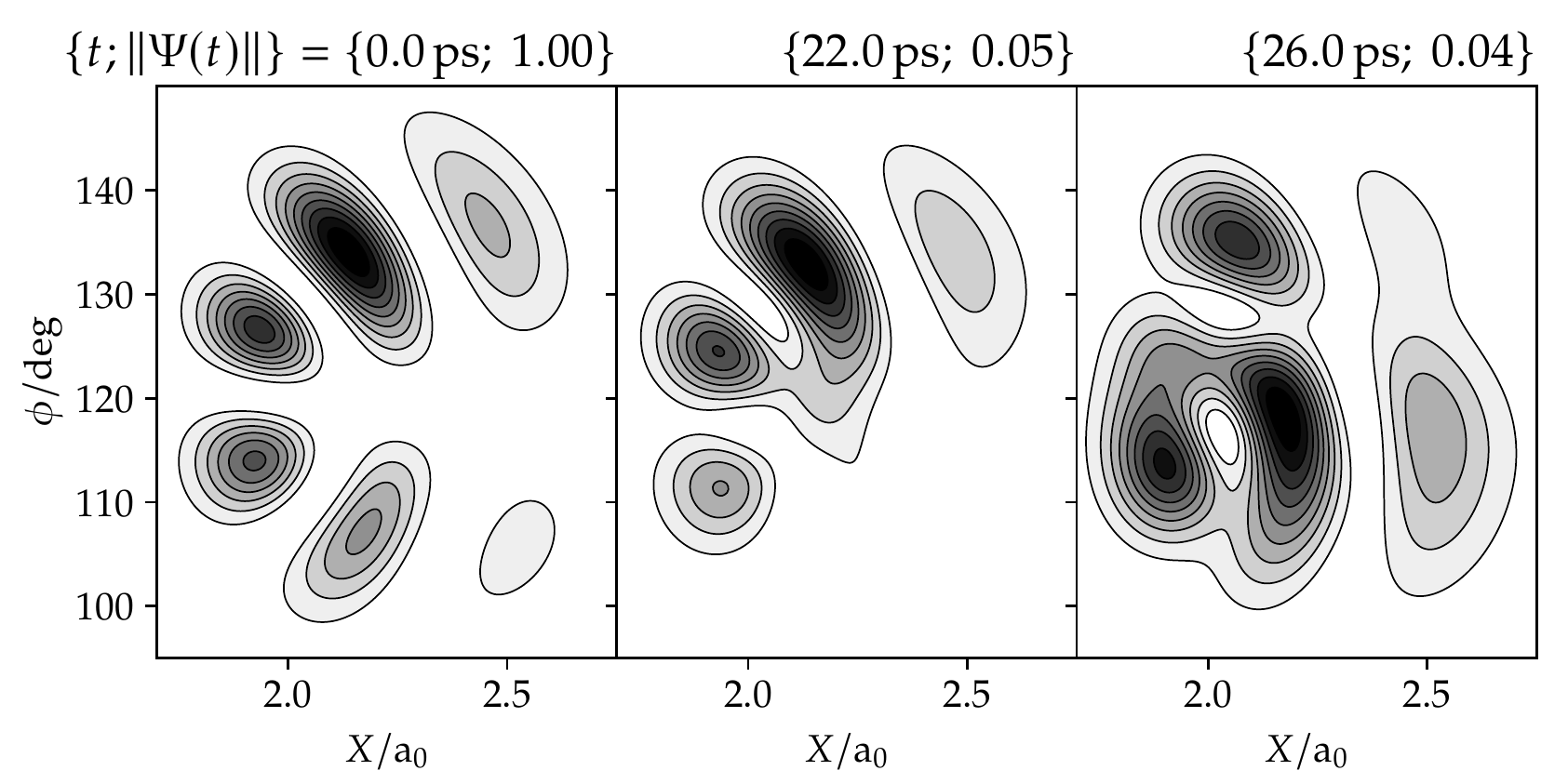}
 \caption{Same as \autoref{fig:decay_034} but showing cuts of $|\Psi(X,Y,\phi; t)|^2$ 
at 
$Y=\unit[2.5]{a_0}$ for the (2,3,1) resonance.}
 \label{fig:decay_231_R}
\end{figure}

A decrease in $v_3$ to $0$ while the other quantum numbers are approximately conserved is 
clearly seen also in state (2,3,1), see \autoref{fig:decay_231} and 
\autoref{fig:decay_231_R}.
Both cuts in $Y$ and $X$ show a decrease in $v_3$ whereas the $v_2=3$ and $v_1=2$ 
components are constant. At $\unit[26]{ps}$, the cut shown in $Y$ (right panel in 
\autoref{fig:decay_231_R}) is more difficult to analyze because contributions from other 
zero-order states become more visible.

The corresponding state on the SAG surface shows no significant IVR. There, the state has 
similar zero-order components except for a reduced $v_1=2$ component. This decreases 
the decay width which is smaller, compared to experiment; see \autoref{tab:energies_SAG}).
The hindered IVR due to the smaller decay width ($\Gamma=\unit[0.36]{cm^{-1}}$) is in 
agreement with the discussion in \autoref{sec:mech_5}.


\begin{figure}
 \includegraphics[width=\columnwidth]{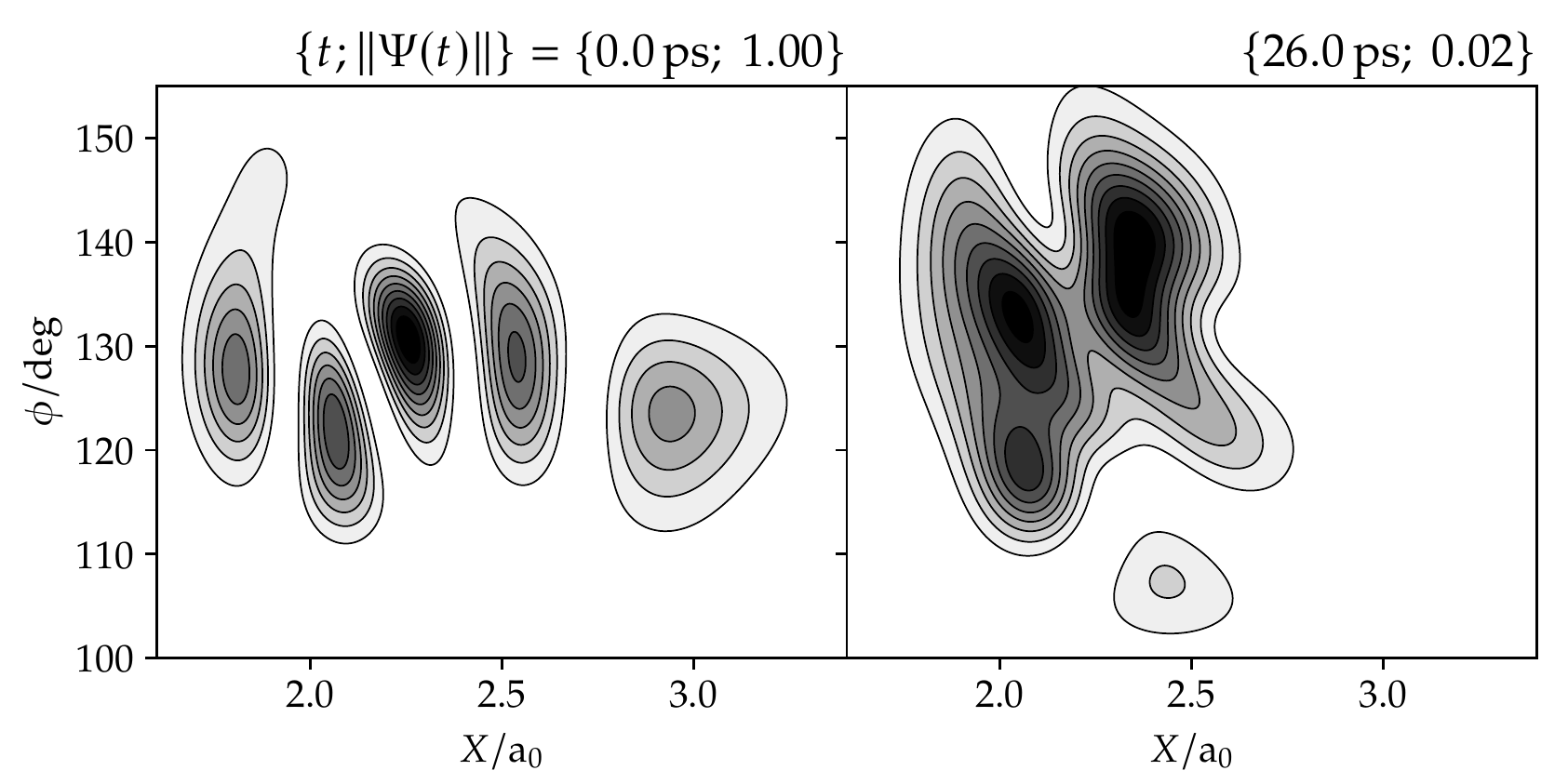}
 \caption{Same as \autoref{fig:decay_034} but showing cuts of $|\Psi(X,Y,\phi; t)|^2$ 
at $Y=\unit[2.09]{a_0}$ for the (1,4,1) resonance.}
 \label{fig:decay_141_R}
\end{figure}

As (2,3,1), state (1,4,1) decays in $v_3$. Here, the $v_2=5$ component (see 
\autoref{tab:energies}) becomes more pronounced. 
The contribution of the (4,2,0) zero-order state shows a decay from $v_1=4$ to $v_1=1$, 
see 
\autoref{fig:decay_141_R}. There, the wavefunction looks like it has increased in $v_3$ 
but an inspection of the phase cannot clearly confirm this.

State (1,4,1) computed on the SAG surface shows a similar initial state (with a reduced 
component of the (2,3,1) state) and a similar decay mechanism. 
There, $\Gamma$ is lower ($\unit[0.85]{cm^{-1}}$ compared to the experimental value of 
$\unit[6]{cm^{-1}}$; see \autoref{tab:energies_SAG}), but not too low for IVR, compared 
to the resonances that show no pronounced IVR.

A sequential decay from $v_1=3$ \emph{via} $v_1=2$ to $v_1=\{1,0\}$ 
in conjunction with 
a decay of $v_3=1$ to $v_3=0$ can be observed for resonance (3,2,1).
Again, $v_2$ remains conserved.
Note that this resonance in polyad 5.5 has the highest \ce{D-C} stretch excitation.
Due to the large decay width of $\Gamma=\unit[17]{cm^{-1}}$, the IVR begins quickly and 
there are intermediate states with contributions from many zero-order states that are 
hard to analyze. 
Note that this state has some contributions of $v_3=5$ and, due to this high excitation 
in the bending mode, contributions at linear geometries.
However, they are less pronounced than for the other states with similar contributions.

While a decrease in bending quantum number $v_3$ can be seen in all resonances 
where a clear IVR occurs, a decrease in \ce{D-C} stretch quantum number $v_1$ can only be 
seen for $v_1>2$, namely in states (3,2,1), (1,4,1) and (0,3,4) where the latter two have 
zero-order states with $v_1=4$.

\section{Conclusions}
\label{sec:conclusion}

To summarize, we have employed our newly developed dynamically pruned DVR (DP-DVR) for 
computing resonance states in \ce{DCO} of polyads $5$ and $5.5$  using filter 
diagonalization and for subsequential propagation to analyze their decay mechanisms. 
For selected resonances, the kinetic energy release (KER) spectra were compared to 
experimental 
results obtained from velocity-map images.

For this challenging test case where the wavefunction has to be propagated in the 
asymptotic region for up to 180 picoseconds, DP-DVR works well and the computed resonance 
energies and widths show good agreement with results from the literature.
The computed KER spectra are in good agreement with the experimental spectra.
Two PES have been compared. For many states, the WKS surface\cite{HCO_I_schinke} gives 
slightly better results than the SAG surface.\cite{HCO_pes_song_2013}
Since no PES was constructed with this particular application in mind (decay dynamics of
energetically high-lying resonances far out into the asymptote), 
no quantitative agreement between
theoretical and experimental results can be expected.
Also, the importance of including $J>0$ and Renner-Teller couplings to higher
electronic states at linear geometries still has to be elucidated.

The rovibrational distribution of the asymptotic \ce{CO} states shows major contributions 
only for \ce{C\bond{3}O} stretch quantum numbers $v=0$ and $1$. The distribution in the
\ce{CO} rotational quantum number $j$ is multimodal. For states with larger decay widths 
of $\Gamma > \unit[0.6]{cm^{-1}}$, the initial state already shows the qualitatively 
correct shape of the distributions.
Due to a coupling of the rovibrational states even in the near-asymptotic region, states 
with smaller decay rates show qualitatively different distributions after long 
propagation times.

Analyses of the decay processes confirm the results from a polyad model 
Hamiltonian.\cite{DCO_polyad_troellsch_2001,DCO_polyad_troellsch_2001} 
For the studied resonance states, the $v_2$ quantum 
number typically is a conserved quantity while 
the wavefunction part in the interaction region shows a 
decrease in quantum number $v_3$. For high \ce{D-CO} stretch excitations with 
corresponding quantum number $v_2 > 2$, a decay in this quantum number is possible as 
well. In contrast to the polyad model, all resonance {states} show 
significant mixing of zero-order states.

Some {zero-order states}, in particular in polyad $5$, show very strong 
coupling with each other. Almost all zero-order components needed for an IVR are already 
contained in the initial {resonance} state and, for these states, no 
appearance or disappearance of 
additional zero-order components can be observed. 

The qualitative agreements between a polyad model and full quantum dynamics confirm that
there are analyzable residues of orderly IVR mechanisms present in
the strongly anharmonic \ce{DCO} system. However, not surprisingly, our full
quantum dynamics also indicate where and how a mechanistic understanding
has to transcend a zero-order model picture.

\section*{Supplementary material}
See supplementary material for plots of the PES, the propagation times, details on the 
diabatization and the asymptotic distributions of all resonance {states}. 

\begin{acknowledgments}
We thank R.~Schinke, L.~Song, A.~van der Avoird and G. C.~Groeneboom for providing us with 
their PES.
H.~R.~L.~thanks R.~Welsch for helpful discussions.
H.~R.~L.~acknowledges support by the Studienstiftung des deutschen Volkes.
J.~R., J.~W.~and F.~T.~gratefully acknowledge the financial support of the 
experimental part of this work by the Deutsche Forschungsgemeinschaft.
J.~R.~is indebted to F.~Renth for discussions regarding the analysis of the photofragment 
images. J.~W.~thanks the Alexander von Humboldt-Stiftung for a research fellowship.
\end{acknowledgments}

\end{document}